\documentclass[preprint,1p,times]{elsarticle}

\biboptions{sort&compress}
\usepackage[numbers]{natbib}
\usepackage{graphicx}
\usepackage{amssymb}
\usepackage{caption}
\usepackage[mathlines,switch]{lineno}
\usepackage{algorithm}
\usepackage{algorithmic}
\usepackage{xcolor}
\usepackage{booktabs}
\usepackage{tabularx}
\usepackage{multirow}
\usepackage{subcaption}
\usepackage{amsmath}
\usepackage{parskip}
\usepackage{tikz}
\usepackage{float}
\usetikzlibrary{arrows.meta, calc, positioning, fit, shapes.geometric}

\usepackage[hidelinks]{hyperref}
\usepackage{cleveref}
\usepackage{makecell}

\crefname{figure}{Fig.}{Figs.}
\Crefname{figure}{Fig.}{Figs.}
\crefname{equation}{Eq.}{Eqs.}
\Crefname{equation}{Eq.}{Eqs.}

\begin{document}
% \modulolinenumbers[3]
\begin{frontmatter}



\title{Rigid-Deformation Decomposition AI Framework for 3D Spatio-Temporal Prediction of Vehicle Collision Dynamics}

%% Author names and affiliations
\author[label1]{Sanghyuk Kim}
\author[label2]{Minsik Seo}
\author[label3]{Sunwoong Yang\corref{cor1}}
\author[label2,label3]{Namwoo Kang\corref{cor2}}

\cortext[cor1]{Corresponding author. email: sunwoongy@kaist.ac.kr}
\cortext[cor2]{Corresponding author. email: nwkang@kaist.ac.kr}

\affiliation[label1]{organization={Department of Mechanical Engineering, KAIST},
            city={Daejeon},
            postcode={34141},
            country={Republic of Korea}}
\affiliation[label2]{organization={Narnia Labs},
            city={Daejeon},
            postcode={34051},
            country={Republic of Korea}}
\affiliation[label3]{organization={Cho Chun Shik Graduate School of Mobility, KAIST},
            city={Daejeon},
            postcode={34051},
            country={Republic of Korea}}            

%% Abstract
\begin{abstract}
    This study presents a rigid-deformation decomposition framework for vehicle collision dynamics that mitigates the spectral bias of implicit neural representations, that is, coordinate-based neural networks that directly map spatio-temporal coordinates to physical fields. We introduce a hierarchical architecture that decouples global rigid-body motion from local deformation using two scale-specific networks, denoted as \textit{RigidNet} and \textit{DeformationNet}. To enforce kinematic separation between the two components, we adopt a frozen-anchor training strategy combined with a quaternion-incremental scheme. This strategy alleviates the kinematic instability observed in joint training and yields a $29.8\%$ reduction in rigid-body motion error compared with conventional direct prediction schemes. The stable rigid-body anchor improves the resolution of high-frequency structural buckling, which leads to a $17.2\%$ reduction in the total interpolation error. Loss landscape analysis indicates that the decomposition smooths the optimization surface, which enhances robustness to distribution shifts in angular extrapolation and yields a $46.6\%$ reduction in error. To assess physical validity beyond numerical accuracy, we benchmark the decomposed components against an oracle model that represents an upper bound on performance. The proposed framework recovers $92\%$ of the directional correlation between rigid and deformation components and $96\%$ of the spatial deformation localization accuracy relative to the oracle, while tracking the temporal energy dynamics with an $8\,\text{ms}$ delay. These results demonstrate that rigid-deformation decomposition enables accurate and physically interpretable predictions for nonlinear collision dynamics.
\end{abstract}

% \centering
% \includegraphics[width=\textwidth]{./Figure/graphical_abstract.pdf}

% \begin{highlights}
% \item Proposed framework decouples rigid-body motion and local deformation
% \item Frozen-anchor training eliminates kinematic coupling and instability
% \item Decomposition smooths the optimization landscape, improving robustness
% \item Framework ensures directional, spatial, and temporal consistency
% \end{highlights}

%% Keywords

\end{frontmatter}


% \linenumbers % Start line numbering
\section{Introduction}
\label{sec:introduction}
Structural crashworthiness is a key requirement in vehicle design. Full-scale crash experiments provide validation but are destructive and costly, which limits design iteration~\citep{baguley2008cost}. As a result, finite element method (FEM) simulations are widely used. These simulations involve millions of degrees of freedom and complex contact interfaces to capture physical responses~\citep{extensive_validation_2004}. During impact, vehicles undergo plastic deformation within milliseconds. To maintain numerical stability, explicit time integration requires microsecond-scale time steps, resulting in runtimes of several hours or days per scenario. This difficulty arises from multi-scale nonlinear crash dynamics. Global rigid-body motion driven by vehicle inertia occurs at low frequencies~\citep{yaw_motion_1_2008, yaw_motion_2_2011}, while local responses such as buckling and plastic hinging appear at high frequencies~\citep{deformation_1_2014, deformation_3_2017}. Resolving both behaviors demands high spatio-temporal resolution, which limits the use of FEM in iterative design workflows.

Deep learning approaches have been adopted to mitigate computational burdens in crashworthiness analysis. Early methods treated the crash event as a black-box function, predicting scalar metrics for design optimization or accident reconstruction. Applications include reconstructing initial impact parameters from final deformation profiles~\citep{chen2021deep}, estimating impact velocities from guardrail damage~\citep{bruski2023speed}, and approximating objective values for topology optimization~\citep{homsnit2024crashworthiness}. These scalar surrogates fail to capture the structural failure mechanisms and kinematic history required for in-depth analysis. To gain temporal insights, subsequent studies inferred structural parameters from 2D post-impact images or predicted 1D time-series data, such as energy absorption and force-displacement curves~\citep{yang2023transfer}. Recently, the field has shifted toward 3D full-field prediction. Recent efforts focus on statically reconstructing deformation fields from damaged point clouds~\citep{xie2022feed} or predicting the dynamic response of component-level structures, such as floor tunnel sections~\citep{thel2024introducing, thel2025accelerating}. However, literature addressing the temporal evolution of full-scale vehicle dynamics remains scarce, leaving the prediction of full-scale dynamic response as an unaddressed challenge.

Autoregressive architectures serve as the standard modeling approach for capturing 3D spatio-temporal dynamics~\citep{dl_collision_4_kim, dl_collision_5_yang, yang2025data, haghighat2021physics, yu2022gradient, kang2025generative}. Graph neural networks (GNNs) utilize message passing among neighboring nodes to model localized interactions~\citep{mgn_lim_message_passing_2018, mgn_lim_mesh_dependency_2020, gnn_review2024, storm2024microstructure, zhao2025physical}. However, vehicle collision dynamics involve global rigid-body motion. The localized aggregation mechanism of GNNs restricts efficient information propagation across the mesh, hindering the capacity to capture global inertial motion. Similarly, sequence-based models, including recurrent neural networks (RNNs) and Transformers, encode temporal dependencies to track nodal displacements~\citep{lstm1997, transformer2017, zhu2025data, wu2025novel, danoun2024fe}. These models rely on sequential time-stepping, where prediction depends on the previous state. This recursive dependency leads to error accumulation over time. Furthermore, for full-scale vehicle models involving tens of thousands of nodes, treating spatial data as sequential inputs results in the curse of dimensionality, imposing high computational costs and inducing training instability~\citep{rnnlimit2013, fang2025efficient}. These limitations render autoregressive architectures impractical for high-resolution crashworthiness analysis.

Implicit neural representations (INRs) provide a resolution-independent alternative by modeling physical fields as continuous functions of spatio-temporal coordinates~\citep{coordinate2021, sun2023physics}. Typical implementations, including multilayer perceptrons (MLPs) and deep operator networks (DeepONets)~\citep{deeponet_2021}, directly map coordinates to physical quantities. Unlike autoregressive architectures, INRs treat time as an independent input variable, enabling point-wise mapping without recursive dependencies. This formulation eliminates error accumulation and avoids processing high-dimensional temporal sequences, making it scalable for 3D spatio-temporal prediction. However, INRs exhibit spectral bias toward low-frequency components~\citep{spectral_1_2019}. In collision dynamics, global rigid-body motion constitutes a low-frequency signal, while local structural buckling and folding represent high-frequency signals. As a result, standard single INRs yield smoothed predictions that fail to capture structural failure modes.

To address this limitation, we propose a hierarchical deep learning framework based on rigid-deformation decomposition. The framework decouples global rigid-body motion from local structural deformation. \textit{RigidNet} predicts low-frequency global motion and serves as a kinematic anchor, while \textit{DeformationNet} models high-frequency residual deformation. This decoupling allows each network to specialize in a specific spectral band and capture structural details that single-network architectures fail to resolve. The contributions of this study are summarized as follows:
\begin{itemize}
    \item We propose a multi-scale deep learning framework based on rigid-deformation decomposition to address the spectral bias of single-network architectures in crash dynamics prediction. The framework decouples global rigid-body motion and local structural deformation using two specialized networks, \textit{RigidNet} and \textit{DeformationNet}.

    \item We introduce a divide-and-conquer strategy that decouples global rigid-body motion and local structural deformation, reducing the learning complexity of entangled multi-scale crash dynamics compared to single-network architectures.

    \item We assess the physical consistency of the proposed framework through directional alignment between global rigid-body motion and local deformation, spatial localization and temporal evolution of deformation, demonstrating improved robustness over existing approaches.
\end{itemize}

The remainder of this paper is organized as follows. Section~\ref{sec:methodology} details the proposed decomposition framework, outlining the architecture and sequential training strategy. Section~\ref{sec:experimental_design} describes the dataset generation, baseline and oracle models, and evaluation protocol. Section~\ref{sec:performance_evaluation} presents the performance analysis, validating the necessity of the decomposition strategy and examining the impact of architectural choices on accuracy and generalization. Section~\ref{sec:verification} verifies the physical validity of the learned components, confirming the consistency of the predicted dynamics. Finally, Section~\ref{sec:conclusion} summarizes the key findings and suggests directions for future work.

\section{Proposed rigid-deformation decomposition framework}
\label{sec:methodology}
Vehicle collision dynamics are characterized by the superposition of global rigid-body motion and spatially varying structural deformation~\citep{shabana1997flexible}. To decouple these multi-scale behaviors, we adopt a hierarchical decomposition framework employing two scale-specific networks and a sequential training strategy.

\subsection{Problem definition}
\label{subsec:problem_definition}
\begin{figure}[H]
\centering
\includegraphics[width=0.6\linewidth]{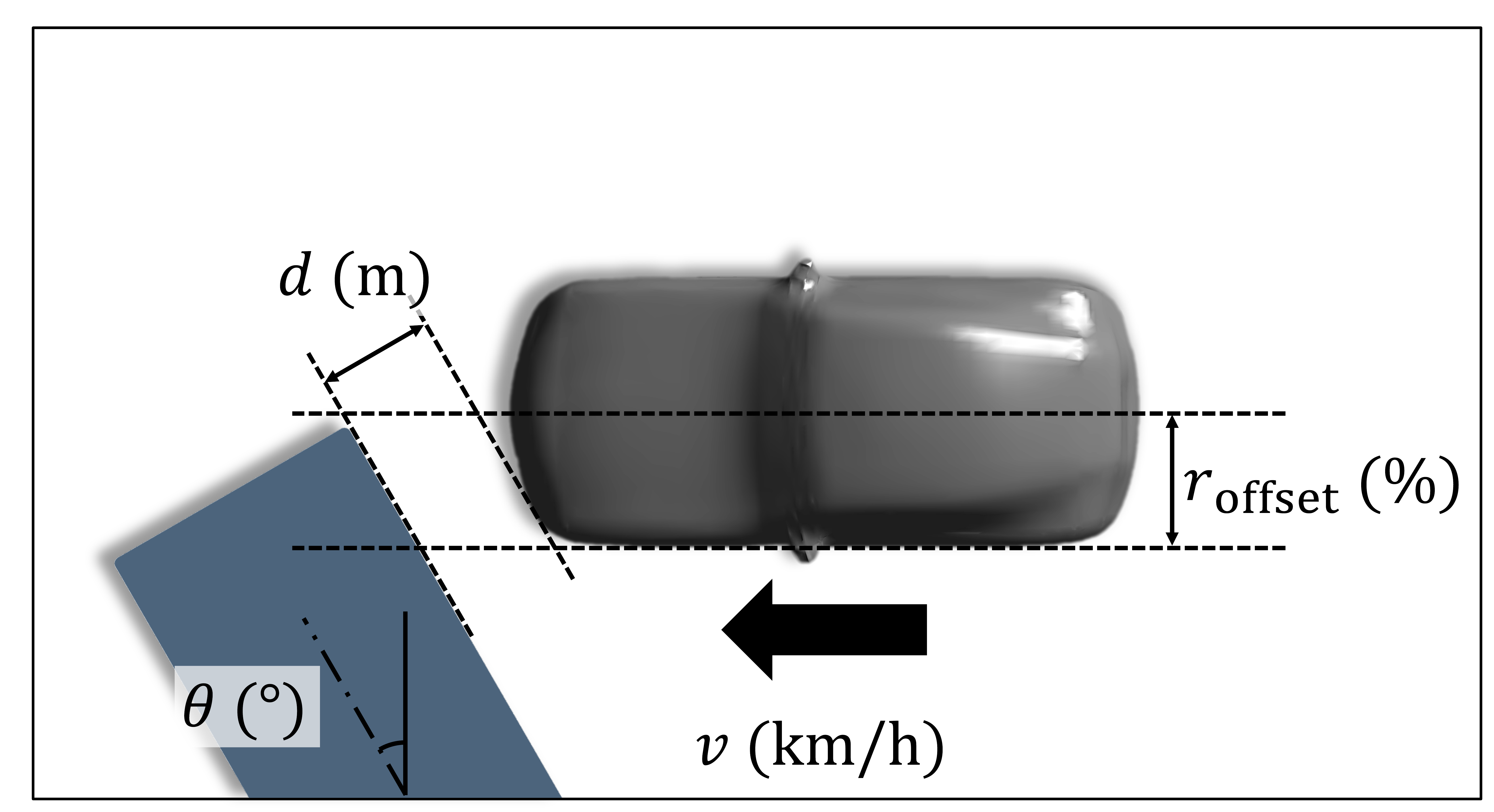}
\caption{
Definition of the collision parameters: initial velocity $v$, collision angle $\theta$, offset ratio $r_{\text{offset}}$, and relative distance $d$.
}
\label{fig:scenario_overview}
\end{figure}

The New Car Assessment Program (NCAP) requires evaluation under diverse frontal, oblique, and offset impact configurations~\citep{ncap_diverse_2016, euroncap2017}. Aligning with these regulatory standards, the collision scenario is parameterized by four variables: the initial vehicle speed $v$ ($\text{km/h}$), the collision angle $\theta$ ($^{\circ}$) between the vehicle heading and the barrier, the frontal offset ratio $r_{\text{offset}}$ (\%), and the relative distance $d$ ($\text{m}$) between the vehicle front and the barrier. These variables form the parameter vector $\boldsymbol{\eta} = [v, \theta, r_{\text{offset}}, d]^{\mathsf{T}},$ defined within the ranges $v \in [40,80]$, $\theta \in [0,45]$, $r_{\text{offset}} \in [5,100]$, and $d \in [0.1,2.0]$. This parameter space covers standard frontal and oblique regulatory tests~\citep{ncap_diverse_2016, euroncap2017, ragland2001evaluation, iihs2012smalloverlap}. Figure~\ref{fig:scenario_overview} illustrates the physical definition of these parameters.

\subsection{Decomposition framework architecture}
\label{subsec:decomposition_architecture}
\begin{figure}[H]
\centering
\includegraphics[width=1\linewidth]{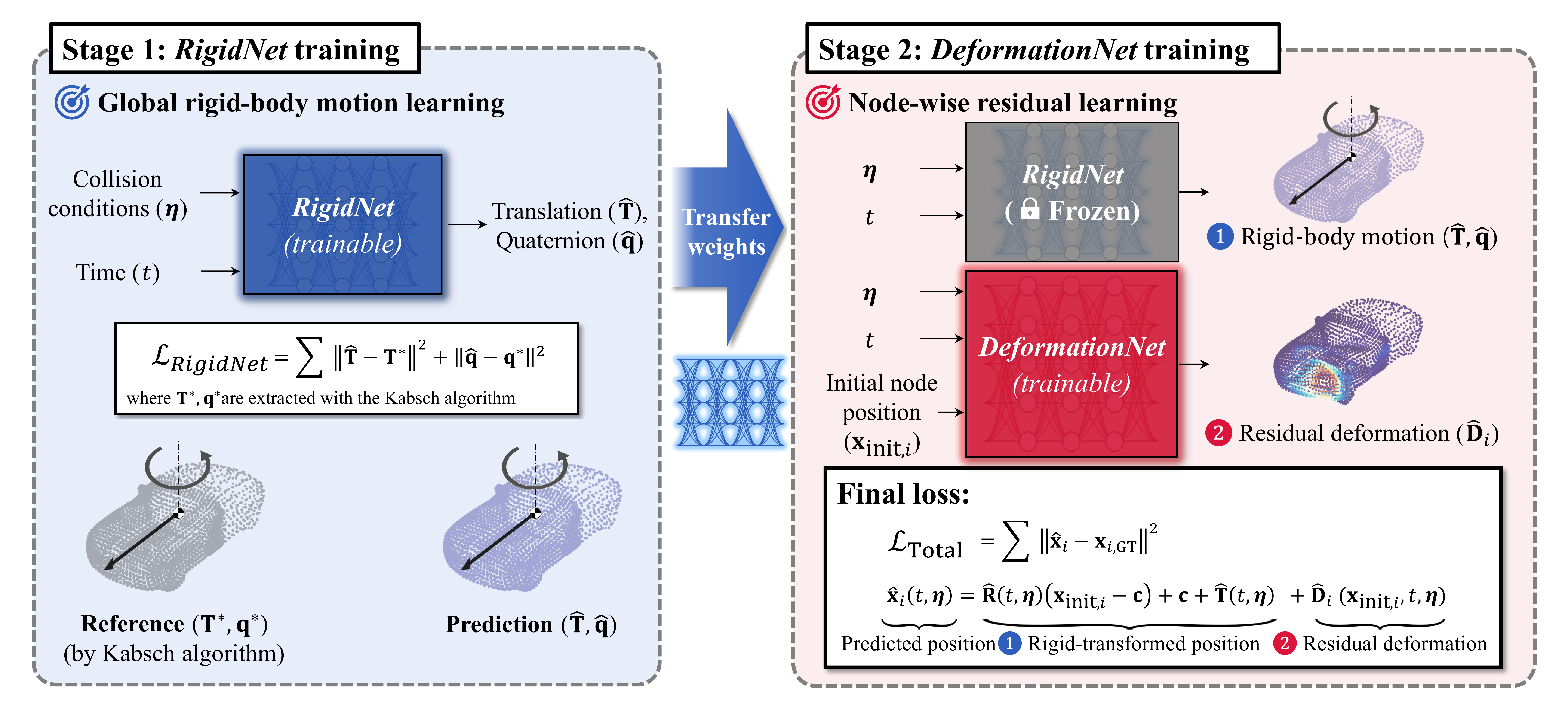}
\caption{Overview of the rigid-deformation decomposition framework. \textit{RigidNet} predicts rigid-body motion, and \textit{DeformationNet} predicts the residual deformation field. A sequential training strategy is employed: Stage~1 trains \textit{RigidNet}, and Stage~2 trains \textit{DeformationNet} with \textit{RigidNet} frozen.}
\label{fig:Architecture}
\end{figure}

Given initial node coordinates $\mathbf{x}_{\text{init},i} \in \mathbb{R}^3$, time $t$, and collision parameters $\boldsymbol{\eta}$, the objective is to predict the spatio-temporal position $\hat{\mathbf{x}}_i$ approximating the finite element solution $\mathbf{x}_{\text{target},i}$. As illustrated in Figure~\ref{fig:Architecture}, the total motion is modeled as the superposition of a global rigid-body transformation and a residual deformation field:
\begin{equation}
\hat{\mathbf{x}}_i(t, \boldsymbol{\eta}) = 
\underbrace{\hat{\mathbf{R}}(t, \boldsymbol{\eta})\!\left( \mathbf{x}_{\text{init},i} - \mathbf{c} \right) + \mathbf{c} + \hat{\mathbf{T}}(t, \boldsymbol{\eta})}_{\text{Global rigid-body motion (\textit{RigidNet})}} + 
\underbrace{\hat{\mathbf{D}}_i(\mathbf{x}_{\text{init},i}, t, \boldsymbol{\eta})}_{\text{Local deformation (\textit{DeformationNet})}},
\label{eq:total_prediction}
\end{equation}
where $\mathbf{c} = \tfrac{1}{N}\sum_{j=1}^N \mathbf{x}_{\text{init},j}$ denotes the centroid of the initial configuration.

\textit{RigidNet} maps $(t,\boldsymbol{\eta})$ to the global rigid-body motion, predicting the translation $\hat{\mathbf{T}} \in \mathbb{R}^3$ and rotation matrix $\hat{\mathbf{R}}$, which is parameterized as a unit quaternion $\hat{\mathbf{q}}$ to ensure stability. \textit{DeformationNet} maps $(\mathbf{x}_{\text{init},i},t,\boldsymbol{\eta})$ to the residual displacement $\hat{\mathbf{D}}_i$, capturing non-rigid distortions. By computing displacement relative to this moving reference frame, the framework decouples the global trajectory from the local structural response, reducing the complexity of the learning task for each network.
\begin{table}[H]
\centering
\caption{Specifications of \textit{RigidNet} and \textit{DeformationNet}. Values in parentheses denote dimensionality.}
\label{tab:architectures}
\begin{tabular}{l l l l}
\hline
\makecell{\textbf{Network}} &
\makecell{\textbf{Input}} &
\makecell{\textbf{Output}} &
\makecell{\textbf{Architecture}} \\ \hline
\textit{RigidNet}
& $t$ (1), $\boldsymbol{\eta}$ (4)
& $\hat{\mathbf{q}}$ (4), $\hat{\mathbf{T}}$ (3)
& 3-layer MLP (256) \\
\textit{DeformationNet}
& $\mathbf{x}_{\mathrm{init}}$ (3), $t$ (1), $\boldsymbol{\eta}$ (4)
& $\hat{\mathbf{D}}_i$ (3)
& 6-layer MLP (256) \\
\hline
\multicolumn{4}{l}{\footnotesize All layers use ReLU activation and He (Kaiming) uniform initialization.}
\end{tabular}
\end{table}

\subsection{Stage 1: supervised pre-training of \textit{RigidNet}}
\label{subsec:stage1_RigidNet}
Stage~1 focuses on the supervised optimization of \textit{RigidNet} to capture global rigid-body dynamics. The ground-truth translation $\mathbf{T}^*$ and rotation matrix $\mathbf{R}^*$ are extracted via the Kabsch algorithm (Section~\ref{subsec:dataset_kabsch}). Here, $\mathbf{R}^*$ is converted to a unit quaternion $\mathbf{q}^*$ to serve as the training target, avoiding the singularity issues inherent to Euler angles (as detailed in Section~\ref{subsubsec:ablation_rigidnet}). Accordingly, \textit{RigidNet} maps the input $(t, \boldsymbol{\eta})$ to the translation $\hat{\mathbf{T}} \in \mathbb{R}^3$ and rotation $\hat{\mathbf{q}} \in \mathbb{R}^4$. The predicted quaternion is normalized to satisfy the unit norm constraint:
\begin{equation}
\mathbf{q} = \frac{\hat{\mathbf{q}}}{\lVert \hat{\mathbf{q}} \rVert}.
\label{eq:quaternion_normalize}
\end{equation}
To improve temporal stability, \textit{RigidNet} learns the incremental motion between consecutive time steps rather than absolute displacement. This approach transforms the long-horizon prediction task into a sequence of local transitions, mitigating error accumulation and preserving temporal continuity~\citep{mgn_2020, dl_collision_4_kim, yang2025model}. The target increments are defined as:
\begin{equation}
\Delta \mathbf{T}^*(t) =
\mathbf{T}^*(t) - \mathbf{T}^*(t-\Delta t),
\qquad
\Delta \mathbf{q}^*(t) =
\mathbf{q}^*(t) \otimes \big(\mathbf{q}^*(t-\Delta t)\big)^{-1},
\label{eq:rigid_increment_def}
\end{equation}
where $\otimes$ denotes quaternion multiplication. The absolute motion is reconstructed through cumulative summation and composition:
\begin{equation}
\mathbf{T}(t) = \sum_{\tau=1}^{t} \Delta \mathbf{T}(\tau),
\qquad
\mathbf{q}(t) = \Delta \mathbf{q}(t) \otimes \cdots \otimes \Delta \mathbf{q}(1).
\label{eq:rigid_increment_accumulate}
\end{equation}
The loss function minimizes the discrepancy in these increments:
\begin{equation}
\mathcal{L}_{\text{RigidNet}}
= \lVert \Delta \hat{\mathbf{T}} - \Delta \mathbf{T}^* \rVert_2^2
+ \min\!\big(
\lVert \Delta \hat{\mathbf{q}} - \Delta \mathbf{q}^* \rVert_2^2,\;
\lVert \Delta \hat{\mathbf{q}} + \Delta \mathbf{q}^* \rVert_2^2
\big).
\label{eq:RigidNet_loss}
\end{equation}
Here, the quaternion term explicitly addresses the double-cover property inherent to 3D rotations~\citep{kuipers1999quaternions}. Upon completion of Stage~1, \textit{RigidNet} is frozen to serve as a consistent reference frame for the subsequent deformation learning in Stage~2.

\subsection{Stage 2: deformation training with fixed \textit{RigidNet}}
\label{subsec:stage2_deformation}
Stage~2 focuses on learning the local structural response by training \textit{DeformationNet} while keeping \textit{RigidNet} frozen. The target residual deformation is isolated by subtracting the rigid-body motion predicted by \textit{RigidNet} from the ground-truth total displacement:
\begin{equation}
\mathbf{D}_{\text{target},i}(t) =
\mathbf{x}_{\text{target},i}(t) -
\big[\hat{\mathbf{R}}(t)(\mathbf{x}_{\text{init},i}-\mathbf{c}) + \mathbf{c}
+ \hat{\mathbf{T}}(t) \big],
\label{eq:target_deformation}
\end{equation}
where $\hat{\mathbf{R}}$ and $\hat{\mathbf{T}}$ denote the global rotation and translation inferred by the pre-trained \textit{RigidNet}. \textit{DeformationNet} then approximates this residual field by minimizing the MSE over all nodes:
\begin{equation}
\mathcal{L}_{\text{DeformationNet}}
= \frac{1}{N} \sum_{i=1}^{N}
\lVert \mathbf{D}_{\text{target},i}(t)
- \hat{\mathbf{D}}_i(t) \rVert_2^2.
\label{eq:deform_loss}
\end{equation}
This decoupled strategy ensures that \textit{DeformationNet} focuses on the high-frequency local dynamics that the global motion model cannot capture.

\section{Experimental setup}
\label{sec:experimental_design}
We validate the proposed framework using a FEM-based simulation dataset representing vehicle collision scenarios. This section details the dataset generation (Section~\ref{subsec:dataset_generation}), rigid-motion label extraction via the Kabsch algorithm (Section~\ref{subsec:dataset_kabsch}), comparative model architectures (Section~\ref{subsec:comparative_models}), and the training and evaluation protocols (Section~\ref{subsec:training_protocol}).

\subsection{Dataset generation: displacement fields with varying collision parameters}
\label{subsec:dataset_generation}
\begin{figure}[H]
\centering
\begin{subfigure}[b]{0.48\textwidth}
    \includegraphics[width=\textwidth]{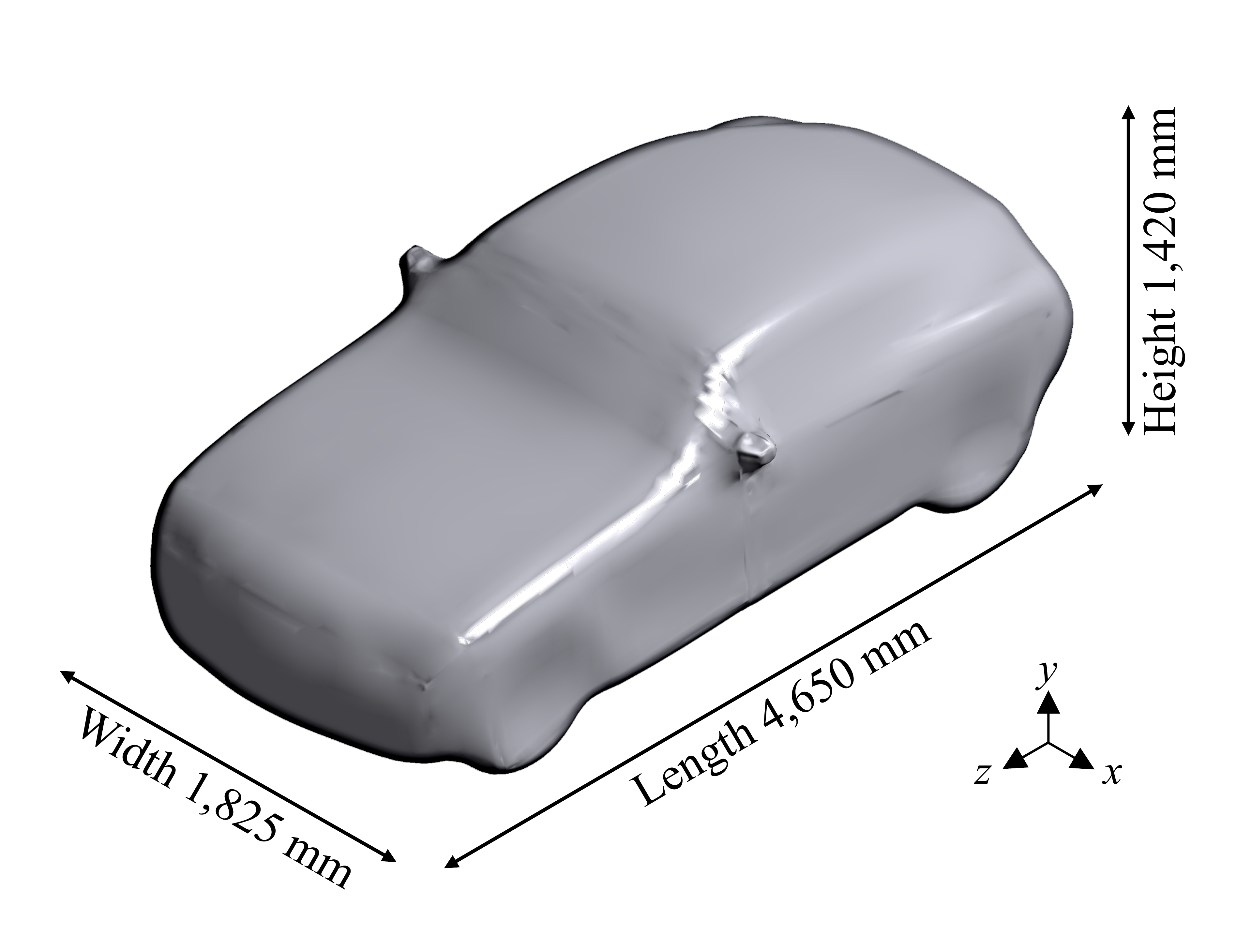}
    \caption{Vehicle geometry}
    \label{fig:vehicle_model_a}
\end{subfigure}
\hfill
\begin{subfigure}[b]{0.48\textwidth}
    \includegraphics[width=\textwidth]{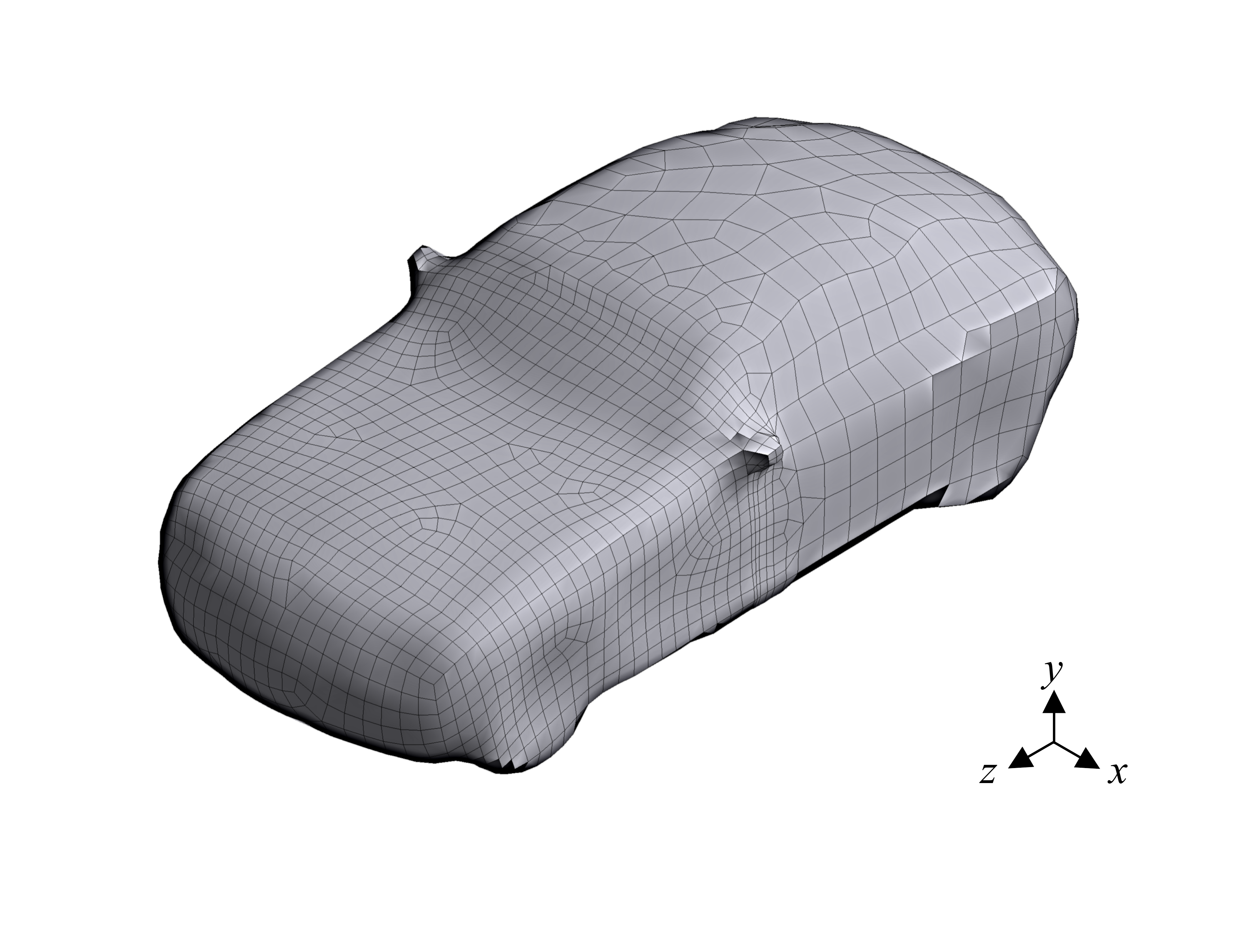}
    \caption{Finite element mesh}
    \label{fig:vehicle_model_b}
\end{subfigure}
\caption{
Vehicle model used in finite element simulations. (a) Geometry scaled to realistic dimensions. (b) Finite element mesh composed of 2,567 nodes and 2,606 elements.
}
\label{fig:vehicle_model}
\end{figure}

\paragraph{Finite element simulation setup}
The vehicle geometry derives from a reference model in the ShapeNet database\footnote{Category: 02958343, Model ID: 1a0bc9ab92c915167ae33d942430658c}~\citep{shapenet2015}, scaled to realistic dimensions as shown in Figure~\ref{fig:vehicle_model_a}. We conducted explicit dynamic simulations using the Ansys AutoDyn solver, discretizing the domain into a finite element mesh comprising 2,567 nodes and 2,606 shell elements with a nominal size of $0.3\,\text{m}$ (Figure~\ref{fig:vehicle_model_b}). The simulation spans $0.4\,\text{s}$, recorded at 100 discrete time steps with an interval of $0.004\,\text{s}$. To capture structural deformation, the vehicle body employs an aluminum alloy modeled with an elasto-plastic constitutive law and bilinear isotropic hardening~\citep{aluminium1994, jones2011structural}. The barrier is defined using the CONC-35MPA concrete model ($35\,\text{MPa}$ compressive strength) to represent specified impact impedance. Contact interactions follow a Coulomb friction model with static and kinetic coefficients of 0.8 and 0.7, respectively~\citep{contact1987}. Comprehensive simulation parameters are provided in \ref{appendix:fem_setup}.

\begin{figure}[H]
\centering
\includegraphics[width=0.6\linewidth]{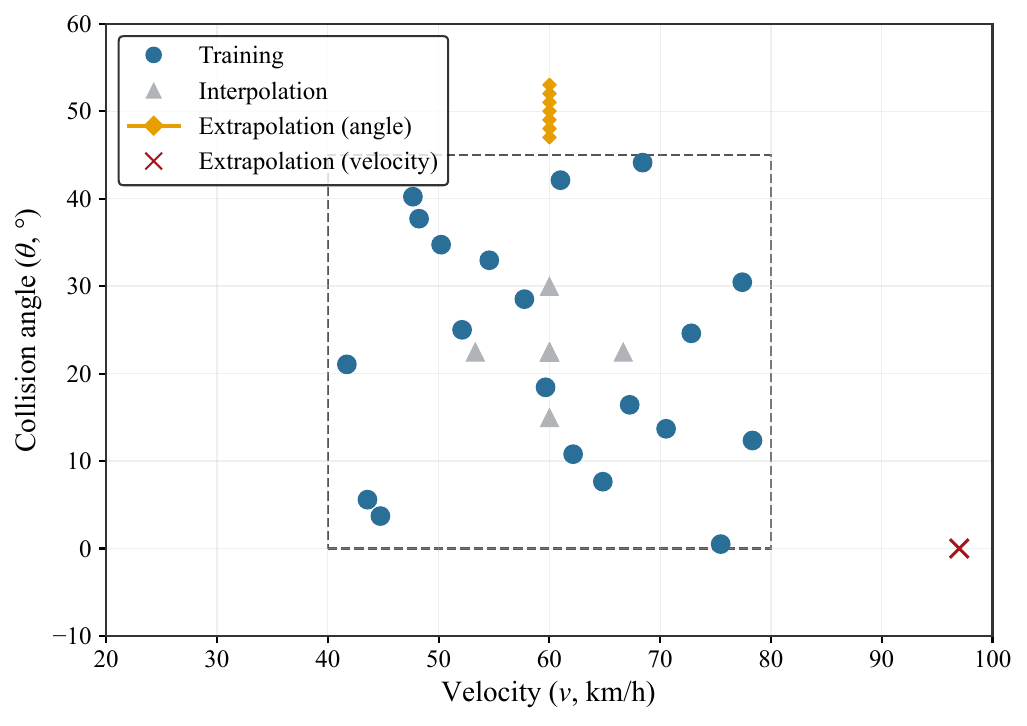}
\caption{
Distribution of collision scenarios in the $v$-$\theta$ parameter space. Blue circular and gray triangular markers denote training and interpolation (validation) scenarios, respectively. Extrapolation (test) scenarios are divided into two types: orange diamond markers connected by solid lines correspond to angular extrapolations (\(46^\circ\!-\!53^\circ\)), and the red cross marker indicates the single velocity extrapolation at \(v = 97\,\text{km/h}\). The dashed rectangle indicates the training domain defined by LHS.
}
\label{fig:dataset_distribution}
\end{figure}

\paragraph{Sampling of collision scenarios}
We generated $20$ uniformly distributed training scenarios using Latin hypercube sampling (LHS)~\citep{lhs2000}. To evaluate generalization capabilities, we defined two distinct evaluation sets: 8 interpolation (validation) scenarios within the training domain, and 9 extrapolation (test) scenarios outside them. The extrapolation set targets extreme parameters, comprising 8 angular deviations ($46^{\circ}\!-\!53^{\circ}$) and a high-velocity impact case ($v = 97\,\text{km/h}$). Figure~\ref{fig:dataset_distribution} visualizes the coverage of these scenarios across the $v$-$\theta$ design space. A sensitivity analysis presented in \ref{appendix:data_sensitivity} confirms that $20$ training samples represent the optimal trade-off between convergence stability and computational efficiency.

\subsection{Rigid-body motion label extraction using the Kabsch algorithm}
\label{subsec:dataset_kabsch}
We extract ground-truth rigid-body motion labels using the Kabsch algorithm~\citep{kabsch1976, kabsch1978}, which computes the optimal rotation and translation minimizing the root mean squared deviation (RMSD) between point sets. Let $N$ denote the total number of nodes, with $\mathbf{x}_{\text{init},i}$ and $\mathbf{x}_{\text{target},i}$ representing the initial and deformed coordinates, respectively. The centroids of the initial ($\mathbf{c}$) and deformed ($\bar{\mathbf{x}}_{\text{target}}$) configurations are computed as:
\begin{equation}
\bar{\mathbf{x}}_{\text{target}}(t,\boldsymbol{\eta})
= \frac{1}{N}\sum_{i=1}^N \mathbf{x}_{\text{target},i}(t,\boldsymbol{\eta}).
\end{equation}
Using the centered coordinates $\mathbf{x}'_{\text{init},i} = \mathbf{x}_{\text{init},i} - \mathbf{c}$ and $\mathbf{x}'_{\text{target},i} = \mathbf{x}_{\text{target},i} - \bar{\mathbf{x}}_{\text{target}}$, we construct the cross-covariance matrix $\mathbf{H}$:
\begin{equation}
\mathbf{H}(t,\boldsymbol{\eta}) 
= \sum_{i=1}^N 
\mathbf{x}'_{\text{init},i}
\bigl(\mathbf{x}'_{\text{target},i}(t,\boldsymbol{\eta})\bigr)^{\mathsf{T}}.
\end{equation}
The optimal rotation matrix $\mathbf{R}^*(t,\boldsymbol{\eta})$ and translation vector $\mathbf{T}^*(t,\boldsymbol{\eta})$ are derived via singular value decomposition (SVD), where $\mathbf{H} = \mathbf{U}\boldsymbol{\Sigma}\mathbf{V}^{\mathsf{T}}$ yields $\mathbf{R}^* = \mathbf{V}\mathbf{U}^{\mathsf{T}}$ and $\mathbf{T}^* = \bar{\mathbf{x}}_{\text{target}} - \mathbf{R}^* \mathbf{c}$.

Since \textit{RigidNet} predicts rotation in the quaternion space to ensure continuity, we convert $\mathbf{R}^*$ into the target unit quaternion $\mathbf{q}^* = [q_0, q_1, q_2, q_3]^{\mathsf{T}}$ using the transformation:
\begin{equation}
\begin{aligned}
q_0 &= \tfrac{1}{2}\sqrt{1 + R^*_{00} + R^*_{11} + R^*_{22}}, \\
q_1 &= \frac{R^*_{21} - R^*_{12}}{4q_0}, \quad
q_2 = \frac{R^*_{02} - R^*_{20}}{4q_0}, \quad
q_3 = \frac{R^*_{10} - R^*_{01}}{4q_0}.
\end{aligned}
\end{equation}
These extracted components $\mathbf{q}^*$ and $\mathbf{T}^*$ serve as the ground-truth supervision targets. Additionally, the residual deformation $\mathbf{D}^*(t,\boldsymbol{\eta})$, obtained by subtracting this rigid transformation from the total displacement, provides the reference for the deformation component used in the oracle analysis (Section~\ref{subsubsec:oracle_models}).

\subsection{Analysis of multi-scale rigid-deformation dynamics}
\label{subsec:multi_scale_dynamics}
\begin{figure}[H]
\centering
\begin{subfigure}[b]{0.48\textwidth}
    \includegraphics[width=\textwidth]{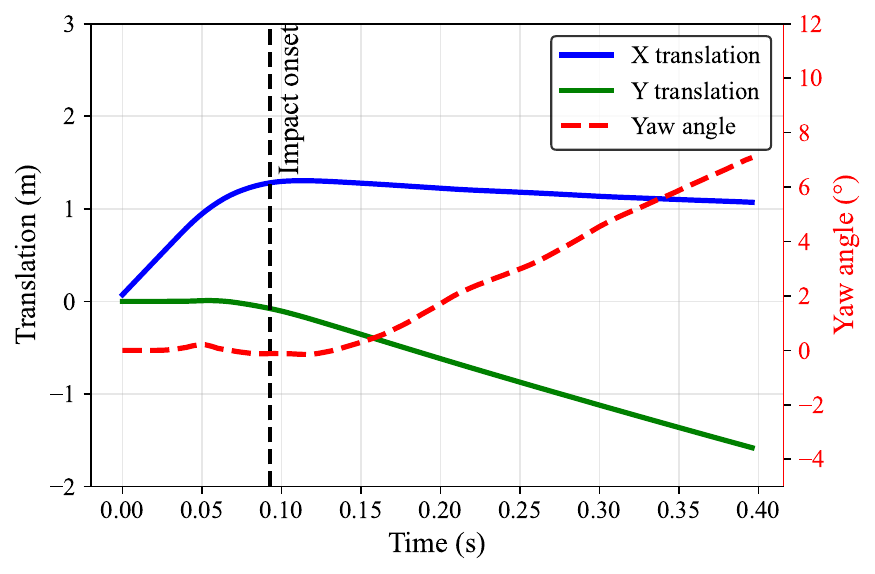}
    \caption{Global rigid-body motion}
    \label{fig:multi_scale_a}
\end{subfigure}
\hfill
\begin{subfigure}[b]{0.48\textwidth}
    \includegraphics[width=\textwidth]{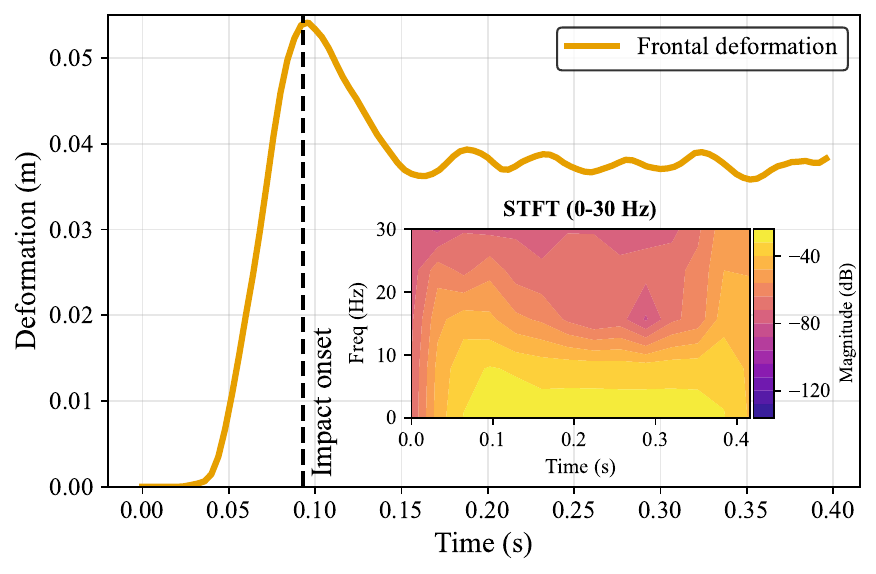}
    \caption{Local deformation with STFT analysis}
    \label{fig:multi_scale_b}
\end{subfigure}
\caption{
Multi-scale dynamics of a vehicle collision. (a) Global rigid-body motion shows smooth, low-frequency evolution. (b) Local deformation exhibits high-frequency transient behavior ($10$--$25~\mathrm{Hz}$), confirming the spectral separation between the two components. The vertical dotted line marks the onset of impact.
}
\label{fig:multi_scale_dynamics}
\end{figure}

To justify the decomposition strategy, we analyzed the spectral characteristics of the collision dynamics. Figure~\ref{fig:multi_scale_dynamics} compares the temporal evolution of the rigid and deformation components for a representative frontal impact scenario ($v = 60\,\text{km/h}$, $\theta = 0^{\circ}$). As shown in Figure~\ref{fig:multi_scale_a}, the rigid-body motion exhibits a smooth, monotonic trajectory in both translation and rotation, characteristic of low-frequency dynamics. In contrast, the frontal region (spanning the first $20\%$ of the vehicle length) undergoes a sharp transient deformation peaking at $0.08\,\text{s}$, followed by an oscillatory rebound, as illustrated in Figure~\ref{fig:multi_scale_b}. Short-time Fourier transform (STFT) analysis quantifies this timescale discrepancy. As shown in the time-frequency spectrogram in Figure~\ref{fig:multi_scale_b}, the deformation energy is concentrated within the $10$--$25~\mathrm{Hz}$ band during the initial impact phase ($0$--$0.1\,\text{s}$), whereas the rigid-body motion energy remains dominant in the low-frequency range (near $0~\mathrm{Hz}$). This distinct spectral separation confirms that global motion and structural deformation are governed by different temporal scales, necessitating an architecture capable of decoupling these multi-scale dynamics.

\subsection{Baseline and comparative model architectures}
\label{subsec:comparative_models}
To validate the decomposition efficacy, we benchmark the proposed framework against two comparative categories: unified models (Section~\ref{subsubsec:unified_models}) and oracle models (Section~\ref{subsubsec:oracle_models}). We selected these INRs because they share the same input domain ($\mathbf{x}_{\text{init}}$, $t$, $\boldsymbol{\eta}$) and resolution-independent inference capability as the proposed method. This design ensures a comparison where performance differences arise solely from the architectural decoupling of dynamics, rather than discrepancies in input information or discretization.

\subsubsection{Unified models}
\label{subsubsec:unified_models}
Unified models (Figure~\ref{fig:unified_models}) serve as the baseline to demonstrate the limitations of single-network architectures. These models map the spatio-temporal inputs directly to the absolute deformed position $\hat{\mathbf{x}}_i(t,\boldsymbol{\eta})$, which forces a single network to resolve both global rigid-body motion and local deformation. In this study, a standard MLP is used as the representative implicit neural representation to isolate the effect of decomposition. The framework is architecture-agnostic, and any coordinate-based model can replace the MLP without altering the decomposition strategy.

\textbf{Coupled MLP:}
This baseline employs an 8-layer MLP with 256 hidden units (approximately $401\text{k}$ parameters). It treats the spatial, temporal, and conditional inputs as a concatenated flat vector, processing them through a shared pathway without distinguishing between global and local dynamics.

\textbf{DeepONet:}
As a representative operator learning architecture, DeepONet consists of a branch network (4 layers, 256 units) encoding the collision parameters $\boldsymbol{\eta}$ and a trunk network (4 layers, 256 units) encoding the spatio-temporal coordinates $(\mathbf{x}_{\text{init},i},\, t)$. The final displacement $\hat{\mathbf{x}}_i$ is computed via the inner product of their latent vectors. Despite its operator-theoretic foundation, this unified variant still targets the total displacement field with a comparable parameter budget (approximately $400\text{k}$), making it subject to the same multi-scale learning difficulties as the coupled MLP.

\subsubsection{Oracle models}
\label{subsubsec:oracle_models}
To dissect the error sources, we introduce oracle models (Figure~\ref{fig:oracle_models}) that establish upper performance bounds by isolating either the rigid or deformation component. For comparison, these models share the exact same architecture (8-layer MLP, approximately $401\text{k}$ parameters) as the unified coupled MLP.

\textbf{Oracle for deformation:}
This variant quantifies the learnability of the local deformation field. It predicts only the residual deformation $\hat{\mathbf{D}}_i$, while the ground-truth rigid-body motion ($\mathbf{R}^*, \mathbf{T}^*$) is injected during reconstruction:
\begin{equation}
\hat{\mathbf{x}}_i(t,\boldsymbol{\eta}) =
\underbrace{
\mathbf{R}^*(t,\boldsymbol{\eta})\,(\mathbf{x}_{\mathrm{init},i}-\mathbf{c}) 
+ \mathbf{c} + \mathbf{T}^*(t,\boldsymbol{\eta})
}_{\text{Ground-truth rigid-body motion}}
+
\underbrace{
\hat{\mathbf{D}}_i(\mathbf{x}_{\mathrm{init},i},t,\boldsymbol{\eta})
}_{\text{Prediction}},
\label{eq:oracle_rigid}
\end{equation}

\textbf{Oracle for rigid-body motion:}
Conversely, this variant assesses the learnability of global motion. It predicts the rigid-body transformation ($\hat{\mathbf{r}}_i$), while using the ground-truth deformation field ($\mathbf{D}^*$) for reconstruction:
\begin{equation}
\hat{\mathbf{x}}_i(t,\boldsymbol{\eta}) =
\underbrace{
\hat{\mathbf{r}}_i(t,\boldsymbol{\eta})
}_{\text{Prediction}}
+
\underbrace{
\mathbf{D}_i^*(\mathbf{x}_{\mathrm{init},i},t,\boldsymbol{\eta})
}_{\text{Ground-truth deformation}},
\label{eq:oracle_deform}
\end{equation}
In both equations, the superscript $*$ denotes ground-truth components extracted in Section~\ref{subsec:dataset_kabsch}. By selectively replacing prediction targets with ground truth, these oracles enable a component-wise decomposition of the total prediction error.

\subsubsection{Proposed framework}
\label{subsubsec:proposed_framework}
\begin{figure}[H]
\centering
\begin{subfigure}[b]{0.32\textwidth}
    \includegraphics[width=\textwidth]{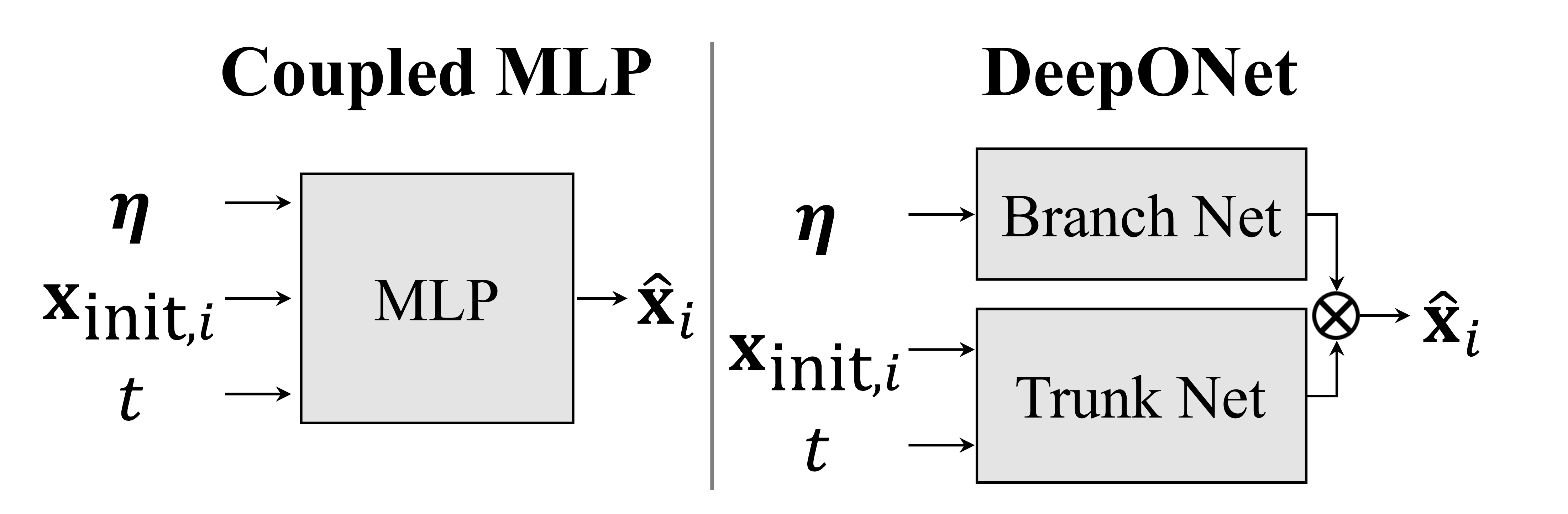}
    \caption{Unified models}
    \label{fig:unified_models}
\end{subfigure}
\hfill
\begin{subfigure}[b]{0.32\textwidth}
    \includegraphics[width=\textwidth]{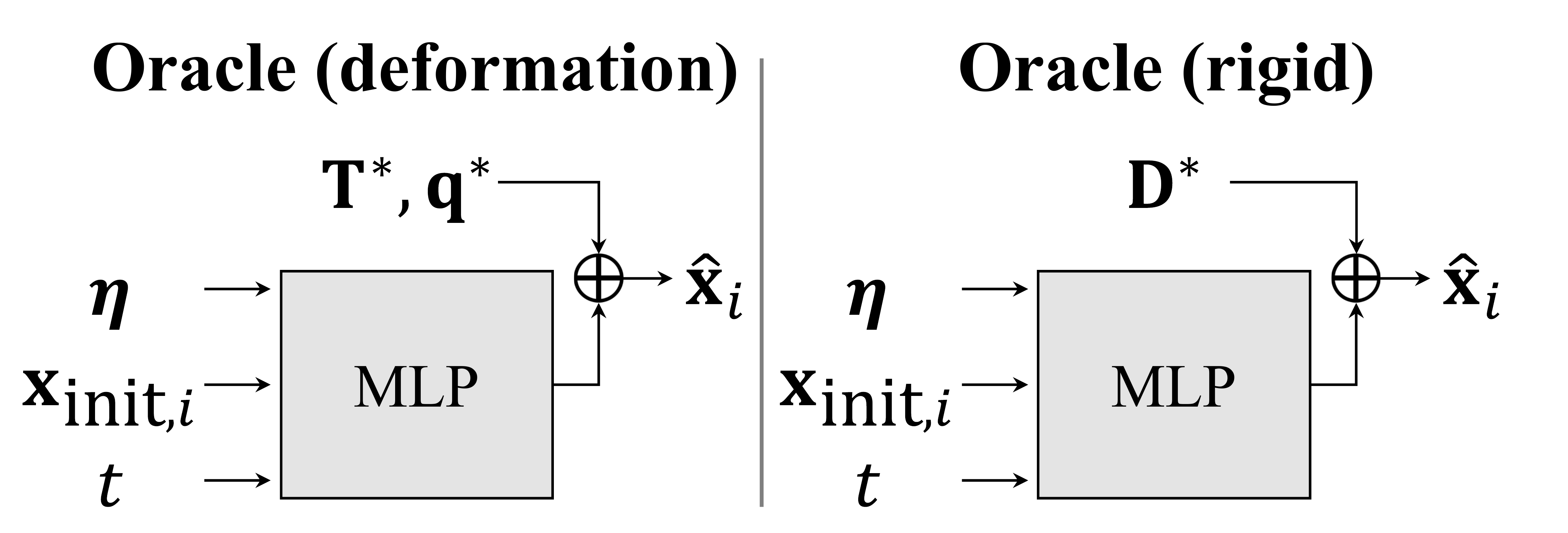}
    \caption{Oracle models}
    \label{fig:oracle_models}
\end{subfigure}
\hfill
\begin{subfigure}[b]{0.32\textwidth}
    \includegraphics[width=\textwidth]{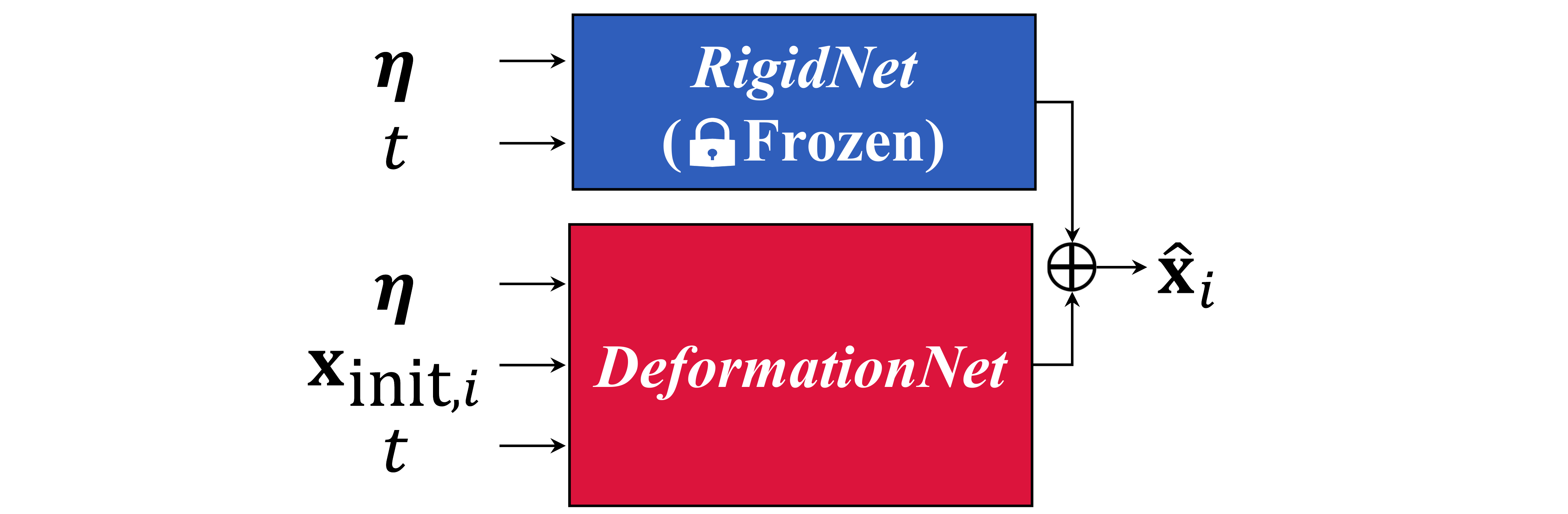}
    \caption{Proposed framework}
    \label{fig:proposed_framework}
\end{subfigure}
\caption{
Comparative architectures. (a) Unified models map inputs directly to the absolute displacement field. (b) Oracle models use ground-truth components ($\mathbf{T}^*, \mathbf{q}^*$ or $\mathbf{D}^*$) to establish performance upper bounds. (c) The proposed framework decomposes dynamics into global and local components, trained via a sequential frozen-anchor strategy to ensure stable residual learning.
}
\label{fig:model_architectures}
\end{figure}

The proposed framework (Figure~\ref{fig:proposed_framework}) implements the kinematic decomposition strategy. It comprises a 3-layer \textit{RigidNet} (256 hidden units, approximately 136k parameters) for global motion prediction and a 6-layer \textit{DeformationNet} (256 hidden units, approximately 268k parameters) for residual deformation. The total parameter count is approximately 404k, deliberately matched to the unified baselines. This controlled parameter budget ensures that any performance improvement is attributed solely to the architectural decomposition and the frozen-anchor training strategy, rather than increased model capacity.

\subsection{Training and evaluation protocol}
\label{subsec:training_protocol}
To enable resolution-independent learning, all models utilize the INR trained on spatio-temporal coordinates. We implemented a random subsampling strategy using only $10\%$ of the full dataset. As validated in \ref{appendix:subsampling}, this sparsity not only reduces computational cost but also acts as a form of regularization, yielding superior generalization performance compared to full-data training.

We performed a hyperparameter search for the unified baselines to identify their performance ceilings. This sweep covered a wide architectural spectrum, ranging from $35\text{k}$ to $2.12\text{M}$ parameters for the coupled MLP and from $68\text{k}$ to $2.12\text{M}$ parameters for DeepONet. This analysis forms the basis for establishing the performance plateau in Section~\ref{subsec:limitations_unified}. However, for the direct architectural comparison, we standardized the model capacity to approximately $400\text{k}$ parameters. All networks were optimized using the Adam algorithm~\citep{adam_2015} on a single NVIDIA RTX A6000 GPU for 200 epochs. The training computational cost was consistent across architectures, with epoch times of approximately $10.63\,\text{s}$ for the coupled MLP, $10.94\,\text{s}$ for DeepONet, and $10.65\,\text{s}$ for the proposed framework. Consequently, the proposed method required approximately $35.50\,\text{min}$ to complete 200 epochs, whereas a single FEM simulation took an average of $1.21\,\text{h}$ on a 96-core processor. The objective function for optimization is the mean squared error (MSE):
\begin{equation}
\text{MSE} = \frac{1}{N T_{\text{steps}}} \sum_{t=1}^{T_{\text{steps}}} \sum_{i=1}^{N} \lVert \hat{\mathbf{x}}_i(t) - \mathbf{x}_{\text{target},i}(t) \rVert_2^2,
\end{equation}
where $N$ and $T_{\text{steps}}$ denote the number of nodes and time steps, respectively. While MSE drives the gradient-based optimization, we report the root mean squared error (RMSE) in Section~\ref{sec:performance_evaluation} to interpret displacement accuracy in physical units.

\section{Performance evaluation and comparative analysis}
\label{sec:performance_evaluation}
This section evaluates the proposed rigid-deformation decomposition framework for vehicle collision dynamics. The analysis is structured to identify the limitations of single-network architectures and validate the structural benefits of our decomposition strategy. Section~\ref{subsec:limitations_unified} examines the limitations of unified models in jointly representing rigid motion and deformation within a single network. Section~\ref{subsec:necessity_decomposition} then compares unified models with oracle models, which establish an upper performance bound, highlighting the necessity of rigid-deformation decomposition. Section~\ref{subsec:ablation_study} performs an ablation study to assess the impact of key architectural choices on model performance. Section~\ref{subsec:loss_landscape} analyzes the optimization landscape, confirming that decomposition facilitates flatter minima associated with improved generalization. Section~\ref{subsec:interpolation} evaluates prediction accuracy for interpolation scenarios. Finally, Section~\ref{subsec:extrapolation} investigates extrapolation capability under configurations beyond the training range.

\subsection{Limitations of unified models}
\label{subsec:limitations_unified}
We evaluate the limitations of unified models in representing both global rigid-body motion and local deformation. Two architectures are considered: a coupled MLP and a DeepONet. Because these models must simultaneously capture low-frequency rigid motion and high-frequency deformation, they are inherently constrained by the spectral bias of INRs, which tend to prioritize smooth, low-frequency components during training~\citep{spectral_1_2019}. To mitigate this limitation, a temporal Gaussian Fourier feature (GFF) encoding~\citep{fourierfeatures2020} is applied to the time input, enabling the model to represent rapid deformation dynamics. The GFF mapping is defined as
\begin{equation}
\gamma(t) = [\cos(2\pi\mathbf{B}t),\, \sin(2\pi\mathbf{B}t)],
\label{eq:gaussian_fourier}
\end{equation}
where \(\mathbf{B} \in \mathbb{R}^{M}\) is sampled from \(\mathcal{N}(0,\sigma^{2})\). The mapping size \(M\) and frequency scale \(\sigma\) control the range and resolution of temporal frequencies represented in the model. A comprehensive hyperparameter search is performed by varying architectural factors (network depth, hidden dimension, latent dimension, activation) and GFF parameters (\(M, \sigma\)). In total, 100 configurations are generated using LHS, as summarized in Table~\ref{tab:hyperparam_space}. This experimental design enables consistent evaluation of how model capacity and temporal encoding influence convergence behavior and prediction accuracy.

\begin{table}[H]
\centering
\caption{Hyperparameter search space for unified models. The DeepONet employs identical configurations for both the branch and trunk networks.}
\label{tab:hyperparam_space}
\begin{tabular}{l r r}
\hline
\makecell{\textbf{Parameter}} &
\makecell{\textbf{Coupled MLP}} &
\makecell{\textbf{DeepONet}} \\ \hline
Network depth      & \makecell[r]{4, 6, 8, 10}      & \makecell[r]{4, 6, 8, 10} \\
Hidden dimension   & \makecell[r]{128, 256, 512}   & \makecell[r]{128, 256, 512} \\
Latent dimension   & \makecell[r]{--}              & \makecell[r]{64, 128, 256} \\
Activation         & \makecell[l]{ReLU, GELU, SiLU} & \makecell[l]{ReLU, GELU, SiLU} \\
$M$                & \makecell[r]{2, 4, 8, 16}     & \makecell[r]{2, 4, 8, 16} \\
$\sigma$           & \makecell[r]{1, 10, 100}      & \makecell[r]{1, 10, 100} \\ \hline
\end{tabular}
\end{table}

\begin{figure}[H]
    \centering
    \begin{subfigure}[b]{0.48\textwidth}
        \includegraphics[width=\textwidth]{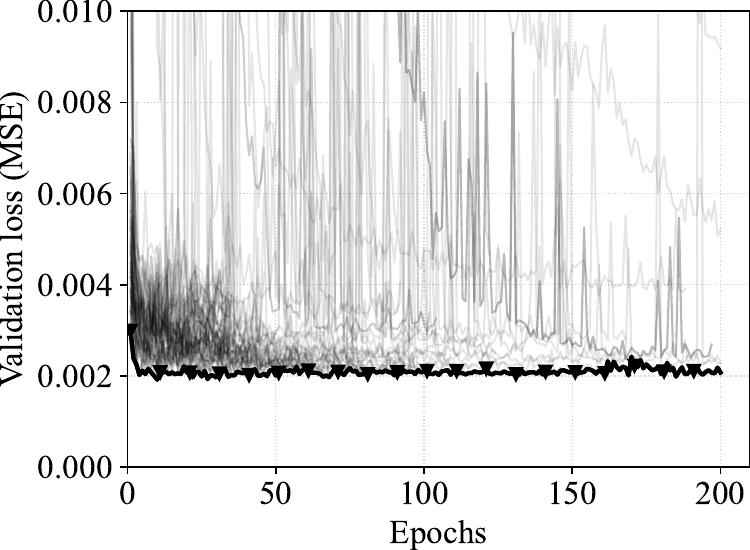}
        \caption{Coupled MLP}
        \label{fig:performance_ceiling_a}
    \end{subfigure}
    \hfill
    \begin{subfigure}[b]{0.48\textwidth}
        \includegraphics[width=\textwidth]{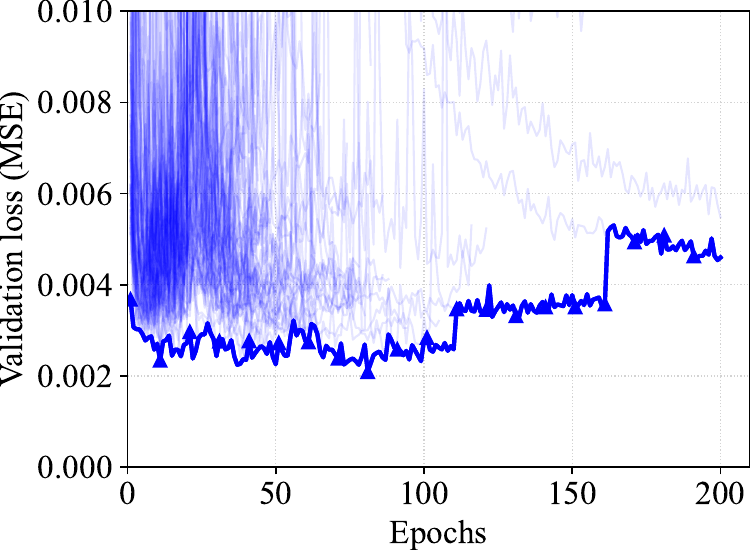}
        \caption{DeepONet}
        \label{fig:performance_ceiling_b}
    \end{subfigure}
    \caption{
        Validation loss of the unified models. The shaded region summarizes 100 runs, and the solid curve represents the minimum validation loss across runs at each epoch. Both models form a similar convergence plateau, indicating that the achievable accuracy is bounded regardless of hyperparameter variation.
    }
    \label{fig:performance_ceiling}
\end{figure}

Figure~\ref{fig:performance_ceiling} shows the validation loss curves of the unified models: coupled MLP (Figure~\ref{fig:performance_ceiling_a}) and DeepONet (Figure~\ref{fig:performance_ceiling_b}). Each model was trained for 200 epochs over 100 distinct hyperparameter configurations. The thin curves correspond to individual training runs, and their spread forms the shaded region representing the overall loss distribution. The solid curve represents the minimum validation loss across all runs at each epoch, defining the best achievable accuracy. Both architectures converge to a similar plateau around \(2.0\times10^{-3}\) despite extensive hyperparameter variation, indicating that their best achievable accuracy is bounded regardless of architectural or hyperparameter configuration. Notably, DeepONet exhibits noticeable instability during training, as observed in Figure~\ref{fig:performance_ceiling_b}. These observations indicate that unified models are constrained by their single-network structure rather than by deficiencies in optimization or model capacity.
Table~\ref{tab:unified_model_stats} summarizes the validation loss statistics over 100 training runs. The coupled MLP and DeepONet show similar convergence trends, and the coupled MLP attains a slightly lower minimum loss. These results are consistent with the convergence plateau in Figure~\ref{fig:performance_ceiling}, indicating that unified models reach an expressive limit that cannot be improved by additional hyperparameter tuning.
\begin{table}[H]
\centering
\caption{
Validation loss statistics from 100 independent trials of unified models.
Lower values indicate better performance.
}
\label{tab:unified_model_stats}

\begin{tabular}{l r r r}
\hline
\makecell{\textbf{Model}} &
\makecell{\textbf{Best (Min)}} &
\makecell{\textbf{Mean $\pm$ Std. Dev.}} &
\makecell{\textbf{Median}} \\ \hline
Coupled MLP & \(2.04\times10^{-3}\) & 0.0310 $\pm$ 0.2616 & \(2.59\times10^{-3}\) \\
DeepONet    & \(2.32\times10^{-3}\) & 0.0711 $\pm$ 0.0082 & \(4.57\times10^{-3}\) \\ \hline
\end{tabular}
\end{table}

\subsection{Necessity of rigid-deformation decomposition}
\label{subsec:necessity_decomposition}
Oracle models isolate the sources of prediction errors, distinguishing between rigid-body motion and local deformation. While the unified model entangles these two components within a single network, oracle models resolve this ambiguity by replacing one motion component with the corresponding ground truth during reconstruction.

\begin{figure}[H]
\centering
\begin{subfigure}[b]{0.48\textwidth}
\includegraphics[width=\textwidth]{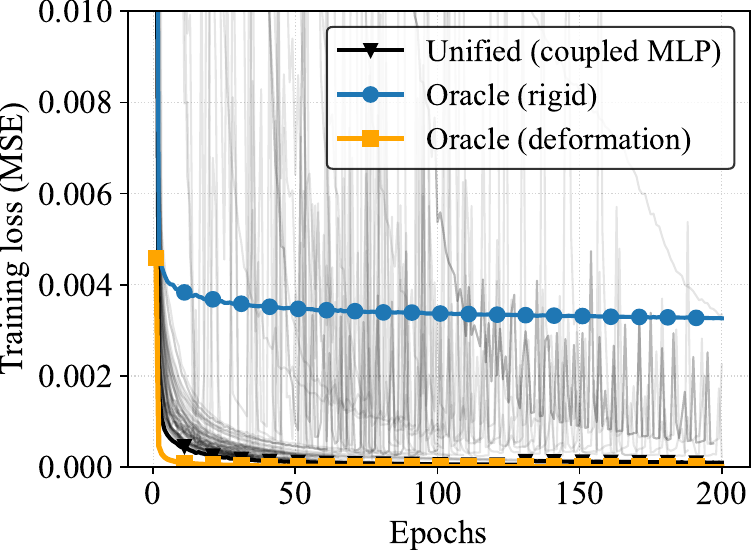}
\caption{Training loss}
\label{fig:necessity_oracle_train}
\end{subfigure}
\hfill
\begin{subfigure}[b]{0.48\textwidth}
\includegraphics[width=\textwidth]{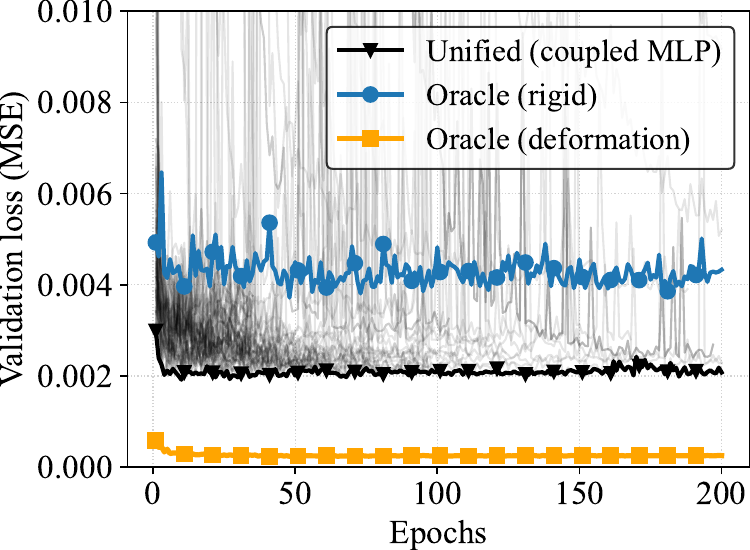}
\caption{Validation loss}
\label{fig:necessity_oracle_val}
\end{subfigure}
\caption{
Training and validation loss curves for the unified model and the oracle models. The oracle for deformation (labeled oracle (deformation)) converges to the lowest loss, whereas the oracle for rigid-body motion (labeled oracle (rigid)) exhibits higher loss than the unified model.
}
\label{fig:necessity_oracle}
\end{figure}

As shown in Figure~\ref{fig:necessity_oracle}, the oracle for deformation exhibits a rapid, stable decrease in training loss (Figure~\ref{fig:necessity_oracle_train}) and achieves a lower validation loss compared with the unified model (Figure~\ref{fig:necessity_oracle_val}). This improvement occurs because eliminating the dominant global rigid-body motion allows the network to focus on the local structural response, reducing the complexity of the mapping task. Consequently, this oracle establishes the upper performance bound for deformation learning. Conversely, the oracle for rigid-body motion yields higher losses than the unified model, even with ground-truth deformation provided. This behavior confirms that rigid-body motion is the dominant source of error. The high loss arises because the INR struggles to approximate a uniform global transformation by independently predicting the displacement of thousands of individual nodes.

Table~\ref{tab:necessity_oracle_stats} quantifies these differences. The oracle for deformation achieves a best validation loss of $2.43\times10^{-4}$, which is approximately eight times lower than the unified model ($2.04\times10^{-3}$). Furthermore, its generalization gap (the difference between best validation loss and corresponding training loss) is also about eight times smaller, indicating the most stable generalization. Conversely, the oracle for rigid-body motion yields a best validation loss more than twice that of the unified model, confirming the rigid-body component as the dominant error source.
\begin{table}[H]
\centering
\caption{
Comparison between the unified (coupled MLP) and oracle models. The oracle for deformation model provides the lowest validation loss and the smallest generalization gap, while the oracle for rigid-body motion model confirms that rigid-body motion dominates the total prediction error.
}
\label{tab:necessity_oracle_stats}
\begin{tabular}{l r r r}
\hline
\makecell{\textbf{Model}} &
\makecell{\textbf{Best validation}} &
\makecell{\textbf{Training@best}} &
\makecell{\textbf{Generalization gap}} \\ \hline
Unified (coupled MLP)          & \(2.04\times10^{-3}\) & \(1.48\times10^{-4}\) & \(1.89\times10^{-3}\) \\
Oracle for rigid-body motion   & \(4.03\times10^{-3}\) & \(3.30\times10^{-3}\) & \(7.24\times10^{-4}\) \\
Oracle for deformation         & \(2.43\times10^{-4}\) & \(8.65\times10^{-6}\) & \(2.34\times10^{-4}\) \\ \hline
\end{tabular}
\end{table}

\subsection{Ablation analysis of the decomposition architecture}
\label{subsec:ablation_study}
We assess the performance gain achieved by the decomposition framework in mitigating the structural and spectral limitations described in Sections~\ref{subsec:limitations_unified} and~\ref{subsec:necessity_decomposition}. The analysis focuses on two aspects: (1) the rigid-body motion representation in \textit{RigidNet} and (2) the training strategy used to integrate \textit{RigidNet} with \textit{DeformationNet}. All results are reported as the mean and standard deviation over five runs (seeds 0--4).

\subsubsection{Rigid-body motion modeling in \textit{RigidNet}}
\label{subsubsec:ablation_rigidnet}
Three \textit{RigidNet} configurations were compared to assess how rotation representation and prediction target affect convergence stability and accuracy:
\begin{itemize}
\item (A) \textbf{Euler-based \textit{RigidNet}}: rotation represented by Euler angles; predicts the absolute rigid transformation at each time step, prone to singularity.
\item (B) \textbf{Quaternion-based \textit{RigidNet}}: rotation represented by a unit quaternion; predicts the absolute rigid transformation at each time step.
\item (C) \textbf{Quaternion-incremental \textit{RigidNet}}: rotation represented as a unit quaternion; predicts temporal increments in rotation and translation, \((\Delta \mathbf{q}, \Delta \mathbf{T})\), between consecutive time steps instead of absolute states.
\end{itemize}

\begin{figure}[H]
\centering
\includegraphics[width=0.8\linewidth]{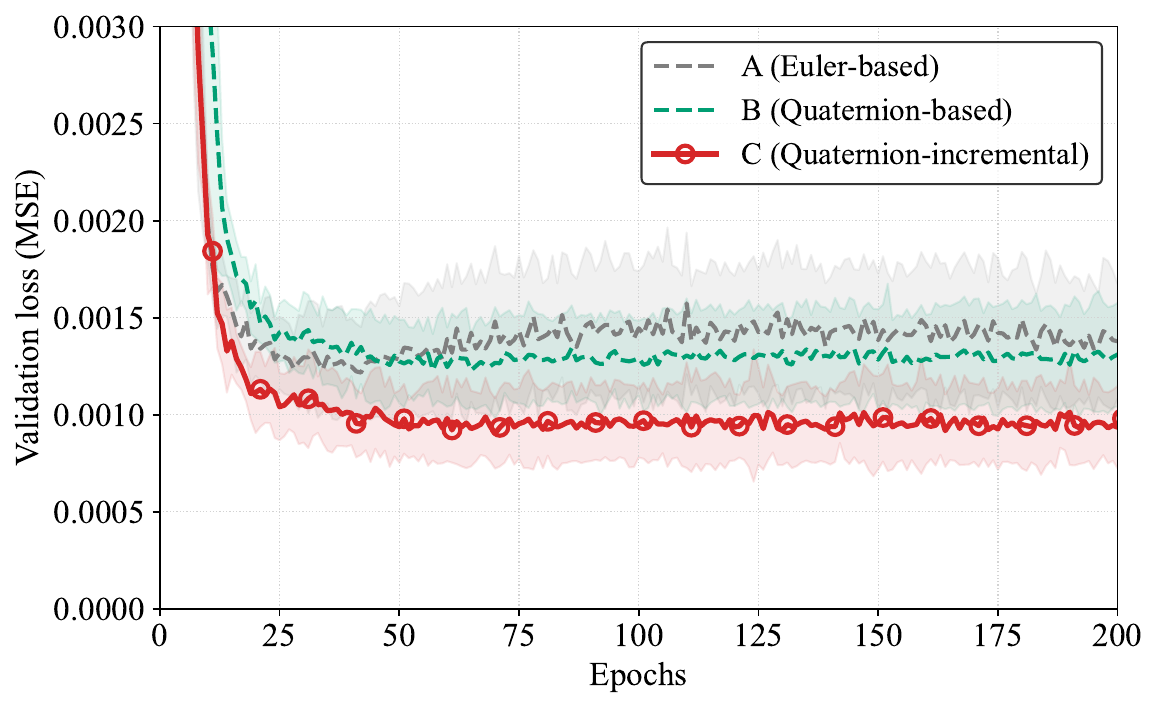}
\caption{
Validation losses for \textit{RigidNet} configurations. Quaternion representation provides smoother loss evolution than Euler angles. The incremental prediction target stabilizes temporal prediction and achieves the lowest validation loss.
}
\label{fig:rigid_ablation}
\end{figure}

The Euler representation, using three angles, introduces discontinuities linked to rotational singularity. In contrast, the quaternion representation, using four normalized components, provides a continuous rotation parameterization that improves numerical smoothness and stability, as shown in Figure~\ref{fig:rigid_ablation}. The incremental model (C) predicts updates relative to the previous time step. This approach constrains the prediction range to local variations, ensuring smaller and smoother updates across time. Physically, this approach aligns with rigid-body motion dynamics, where transformations between consecutive time steps are small and can be assumed linear within short time intervals. Consequently, the model experiences a reduced dynamic range of outputs and enhanced temporal consistency, resulting in more stable training. The results in Table~\ref{tab:ablation} indicate that the quaternion-incremental configuration achieves the lowest mean validation loss of \(0.87 \times 10^{-3}\), representing a $29.8\%$ reduction in error compared to the Euler representation. This improvement arises from achieving physically coherent motion modeling through continuous rotation representation, reduced output range, and enhanced temporal consistency.

\subsubsection{Training strategies for rigid-deformation coupling}
\label{subsubsec:ablation_training_strategies}
Three training strategies were evaluated to assess how the interaction between \textit{RigidNet} and \textit{DeformationNet} affects convergence and stability:
\begin{itemize}
\item (D) \textbf{Baseline, end-to-end}: both networks are trained jointly from scratch.
\item (E) \textbf{Joint fine-tuning}: \textit{RigidNet} is pre-trained, then optimized jointly with \textit{DeformationNet}.
\item (F) \textbf{Proposed, frozen-anchor}: \textit{RigidNet} is pre-trained and fixed, while only \textit{DeformationNet} is updated.
\end{itemize}

\begin{figure}[H]
\centering
\includegraphics[width=0.8\linewidth]{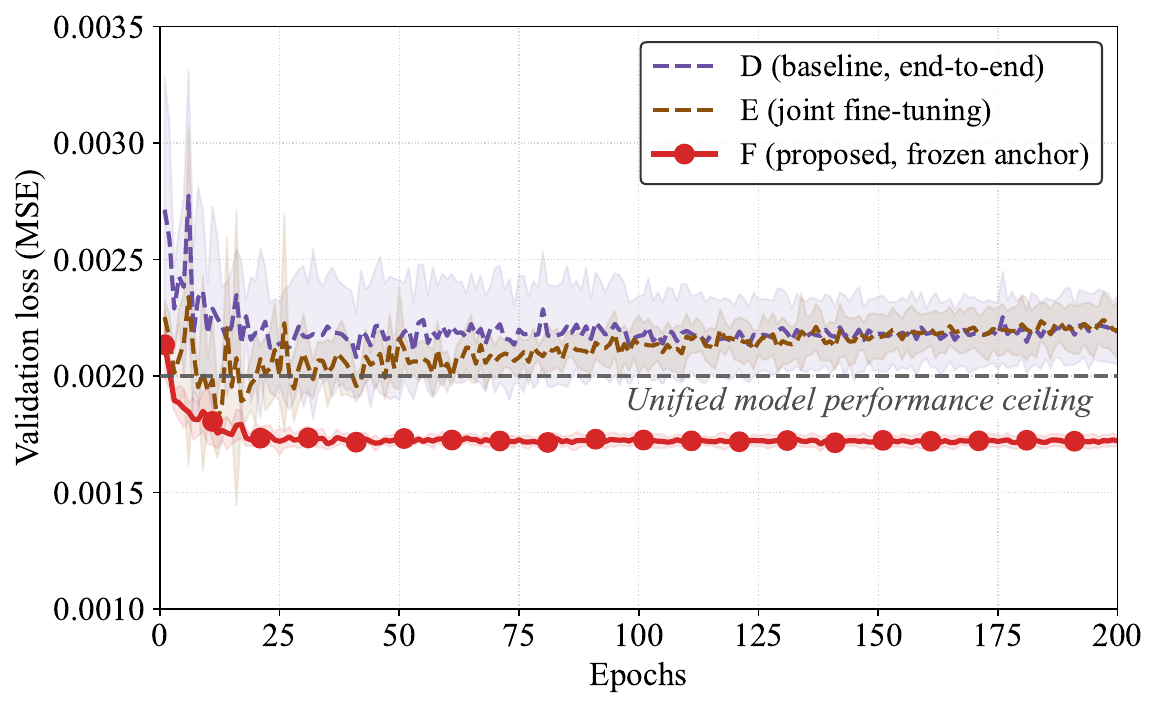}
\caption{
Validation losses for different training strategies. The frozen-anchor configuration (F) converges faster and exhibits lower variance than the end-to-end (D) and joint fine-tuning (E) strategies. The gray dashed line represents the performance ceiling of the unified baseline reported in Table~\ref{tab:unified_model_stats}.
}
\label{fig:deform_ablation}
\end{figure}

Figure~\ref{fig:deform_ablation} compares their validation losses. The frozen-anchor strategy (F) converges faster and exhibits lower variance than (D, E), achieving the lowest validation loss of \(1.69\times10^{-3}\). This performance corresponds to a $17.2\%$ reduction in error compared to the performance ceiling of the unified baseline (\(2.04\times10^{-3}\)) reported in Table~\ref{tab:unified_model_stats}. The superiority of (F) stems from preventing weight interference in the rigid-body prediction during Stage~2 training. In contrast, strategies (D, E) fail to surpass the unified baseline, showing comparable or greater validation losses. This indicates that the decoupling between rigid-body and deformation components is compromised, and \textit{RigidNet} no longer acts as a kinematic anchor. As a result, the coupled training leads to interference between global and local motions, reducing the effectiveness of residual learning. This behavior is further analyzed in Section~\ref{subsec:effect_training_strategies}, where the impact of kinematic divergence on overall convergence is discussed in detail.

Table~\ref{tab:ablation} summarizes the ablation results for the two target dynamics, rigid-body motion and deformation, evaluating rotation strategies for \textit{RigidNet} and training strategies for the proposed framework. Unlike the unified model, temporal encoding yielded minimal benefits within the decomposition framework. Adding GFF reduced the mean validation loss from \(1.69\times10^{-3}\) to \(1.67\times10^{-3}\), a difference within the standard deviation reported in Table~\ref{tab:ablation}. This confirms that modeling rigid-body motion resolves the temporal complexity, rendering additional encoding redundant. Consequently, the final framework relies on the quaternion-incremental scheme and frozen-anchor strategy to ensure convergence and physical interpretability.
\begin{table}[H]
\centering
\caption{
Ablation results. Values are reported as mean~\(\pm\)~standard deviation of the best validation loss. Lower values indicate better performance.
}
\label{tab:ablation}
\begin{tabular}{l l r r}
\hline
\makecell{\textbf{Target dynamics}} &
\makecell{\textbf{Model}} &
\makecell{\textbf{\(\mathrm{MSE}_{\text{rigid}}\,(\times 10^{-3})\)}} &
\makecell{\textbf{\(\mathrm{MSE}_{\text{deform}}\,(\times 10^{-3})\)}} \\ \hline
\multirow{3}{*}{Rigid-body motion}
 & (A) Euler-based             & \(1.24 \pm 0.32\) & -- \\
 & (B) Quaternion-based        & \(1.19 \pm 0.24\) & -- \\
 & \textbf{(C) Quaternion-incremental}  & \(\mathbf{0.87 \pm 0.19}\) & -- \\
\cmidrule(lr){1-4}
\multirow{3}{*}{Deformation}
 & (D) Baseline (end-to-end)   & -- & \(2.08 \pm 0.14\) \\
 & (E) Joint fine-tuning       & -- & \(2.04 \pm 0.10\) \\
 & \textbf{(F) Proposed (frozen-anchor)} & -- & \(\mathbf{1.69 \pm 0.02}\) \\ \hline
\end{tabular}
\end{table}

\subsection{Loss landscape characterization}
\label{subsec:loss_landscape}
To analyze the stability and generalization achieved by the proposed decomposition, we visualized the loss landscapes of the unified, oracle, and proposed frameworks. The loss landscape represents the geometry of the objective function near a converged solution, following the visualization approach of Li et al.~\citep{loss_landscape_viz}. For a model converged with optimal parameters $\boldsymbol{w}^*$, the loss was evaluated over a 2D plane defined by random orthogonal vectors $\boldsymbol{\delta}$ and $\boldsymbol{\xi}$:
\begin{equation}
\boldsymbol{w}(\alpha, \beta) = \boldsymbol{w}^* + \alpha\boldsymbol{\delta} + \beta\boldsymbol{\xi},
\end{equation}
where $\alpha$ and $\beta$ define the coordinates on the visualization plane. The curvature of this surface reflects the generalization tendency of the trained model: networks that converge to wide, flat minima generally exhibit better generalization~\citep{loss_landscape_flat_minima, kim2025projected}, whereas sharp minima with high curvature are associated with overfitting~\citep{loss_landscape_sharp1}.

\begin{figure}[H]
\centering
\includegraphics[width=1.0\linewidth]{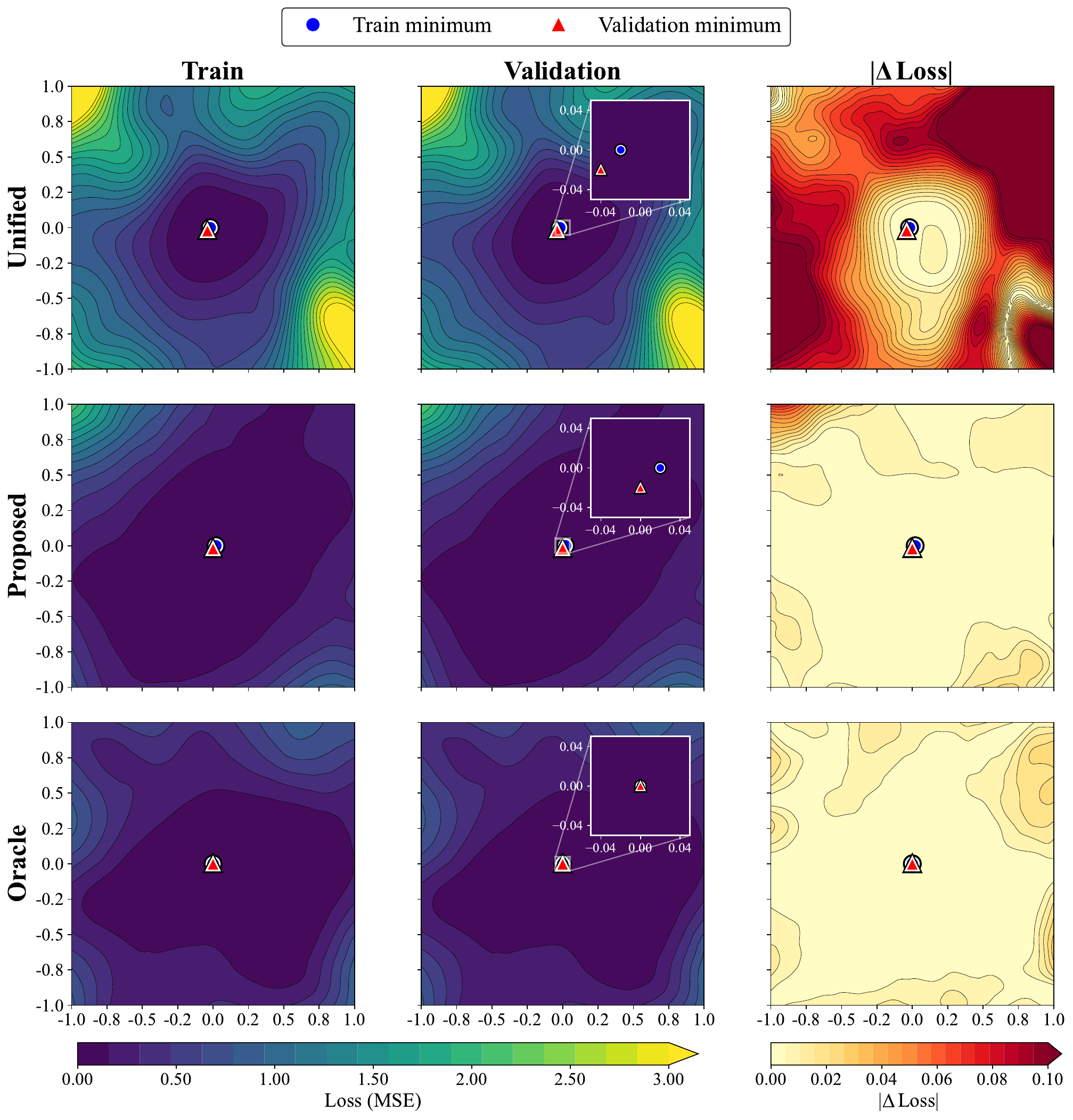}
\caption{
Loss landscape comparison for the unified, proposed, and oracle models. Columns correspond to the training loss, validation loss, and their absolute difference ($|\Delta \text{Loss}|$). Blue circles and red triangles denote the locations of the training and validation minima, respectively. The unified model shows sharp minima with large train-validation separation. The proposed model forms a broad and flat basin with reduced sensitivity. The oracle model also forms a flat basin and achieves the smallest train-validation difference because the rigid-body component is fixed to the ground-truth values.
}
\label{fig:loss_landscape_3x3_grid}
\end{figure}

Figure~\ref{fig:loss_landscape_3x3_grid} compares the 2D loss landscapes of the unified, proposed, and oracle models. The spatial arrangement of the training and validation minima summarizes each model’s response to parameter perturbations. The horizontal distance between the two minima directly indicates how differently the model fits the training and validation datasets. The resulting difference in loss, visualized as $|\Delta\mathrm{Loss}|$, quantifies the generalization gap. The unified model converges to a sharp minimum with high curvature. This sharp landscape increases sensitivity to distribution variations, which enlarges the separation between the minima and produces a large generalization gap. The proposed model forms a flatter basin with a broad low-loss region. The decomposition of rigid-body motion and deformation smooths the optimization surface, which in turn facilitates convergence to a flatter basin. Although the underlying train-validation distribution difference remains comparable to that of the unified model, the flatness of the basin ensures that both minima lie within a low-loss region. This structure produces a small generalization gap and leads to stable generalization performance. The oracle model receives ground-truth rigid-body motion, which removes the dominant source of data distribution shift between the training and validation datasets. As a result, the horizontal separation between the minima becomes minimal. In addition, the simplified optimization task enables convergence to a flat loss surface comparable to that of the proposed model. This flatness mitigates the effect of distribution differences, a capability that the unified model does not possess.

To assess the connectivity and stability of the converged solutions, we trained each model five times using different initialization seeds, resulting in independent parameter solutions $\boldsymbol{w}_1^*,\dots,\boldsymbol{w}_5^*$. A 1D linear interpolation was then performed between all pairwise combinations $(\boldsymbol{w}_i^*, \boldsymbol{w}_j^*)$, where the path between any two solutions is defined as:
\begin{equation}
\boldsymbol{w}(\lambda) = (1-\lambda)\boldsymbol{w}_i^* + \lambda\boldsymbol{w}_j^*, \qquad \lambda\in[0,1].
\end{equation}
Here, $\lambda=0$ corresponds to $\boldsymbol{w}_i^*$ and $\lambda=1$ corresponds to $\boldsymbol{w}_j^*$. We evaluated the loss along this path to quantify the presence of high-loss peaks or connected low-loss regions, which directly indicates the smoothness and stability of the optimization landscape.

\begin{figure}[H]
    \centering
    \includegraphics[width=0.3\linewidth]{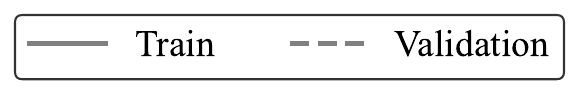}\\[1ex]
    \begin{subfigure}[b]{0.32\linewidth}
        \includegraphics[width=\linewidth]{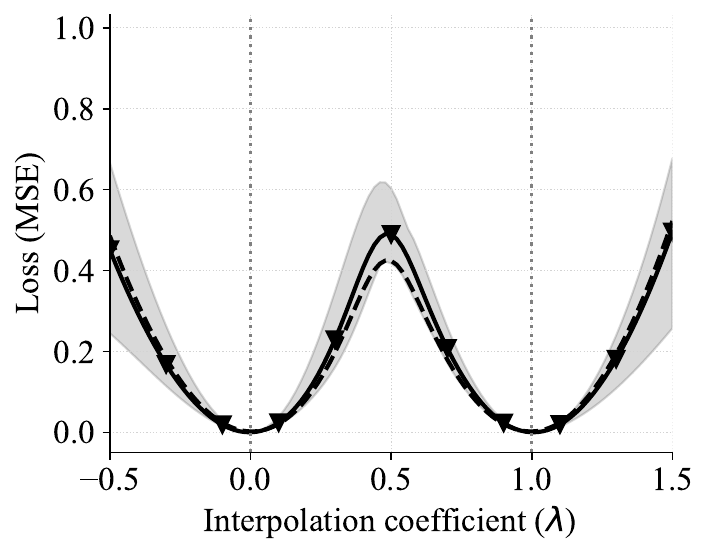}
        \caption{Unified model}
        \label{fig:1D_interpolation_unified}
    \end{subfigure}
    \hfill
    \begin{subfigure}[b]{0.32\linewidth}
        \includegraphics[width=\linewidth]{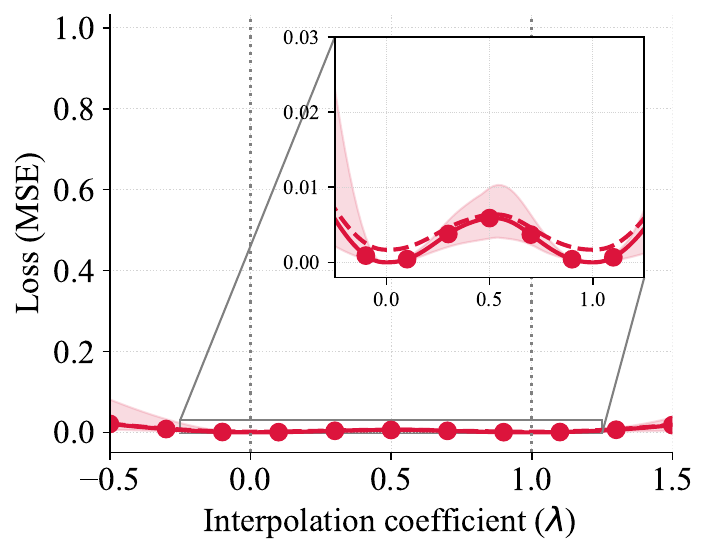}
        \caption{Proposed model}
        \label{fig:1D_interpolation_proposed}
    \end{subfigure}
    \hfill
    \begin{subfigure}[b]{0.32\linewidth}
        \includegraphics[width=\linewidth]{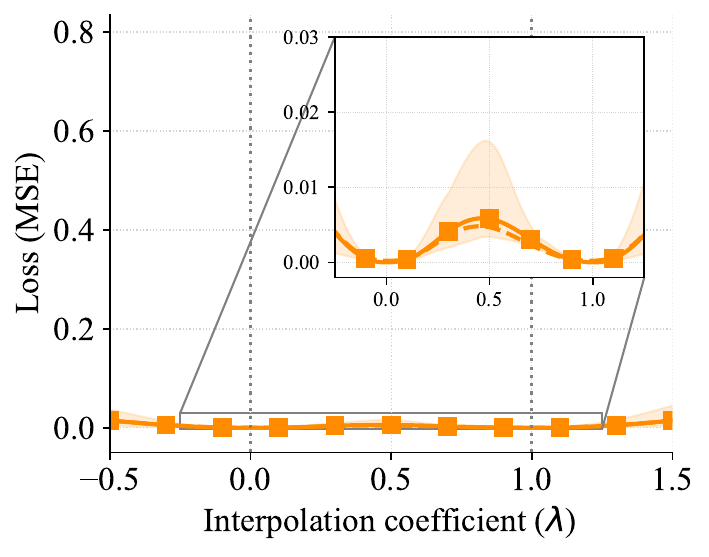}
        \caption{Oracle model}
        \label{fig:1D_interpolation_oracle}
    \end{subfigure}
    \caption{
            1D interpolation of the loss landscapes. Each model was trained with five seeds, and losses were evaluated along interpolation paths between pairwise minima. The unified model exhibits high-loss peaks, whereas the proposed and oracle models form shallow basins with connected low-loss regions.
            }
    \label{fig:1d_loss_interpolation}
\end{figure}

The interpolation results reveal distinct optimization behaviors, as shown in Figure~\ref{fig:1d_loss_interpolation}. The unified model exhibits pronounced loss peaks between minima (Figure~\ref{fig:1D_interpolation_unified}), indicating that independent training runs converge to isolated local solutions separated by steep barriers. This behavior is consistent with a rugged optimization surface. In contrast, the proposed model forms a shallow basin without peaks along the interpolation paths (Figure~\ref{fig:1D_interpolation_proposed}). The loss remains constant, which indicates that the trained solutions lie within a single connected low-loss region. This connectivity suggests that the decomposition of rigid-body motion and deformation reduces interference between global and local dynamics and produces a smoother optimization landscape. The oracle model, due to receiving ground-truth rigid-body motion, inherently reduces optimization complexity and aligns the input space. Its resulting interpolation curves show a similarly shallow basin without intermediate loss peaks (Figure~\ref{fig:1D_interpolation_oracle}). This agreement indicates that the proposed framework achieves the same level of landscape smoothness and solution stability as the oracle model without access to ground-truth rigid-body motion.

\subsection{Evaluation of interpolation in collision scenarios}
\label{subsec:interpolation}
We evaluate interpolation performance using a validation scenario with collision parameters. A representative scenario is analyzed in Figure~\ref{fig:interpolation_field}, defined by $v = 58.21\,\text{km/h}$, $\theta = 22.5^\circ$, $r_{\text{offset}} = 51.35\%$, and $d = 1.07\,\text{m}$. We compute RMSE fields at three key time instants: $t_1 = 0.12\,\text{s}$ (initial impact), $t_2 = 0.20\,\text{s}$ (softening), and $t_3 = 0.36\,\text{s}$ (late impact).

As illustrated in Figure~\ref{fig:interpolation_field}, the unified model exhibits errors stemming from two distinct failure modes. The model fails to localize the high-frequency deformation near the contact zone during the early impact stage ($t_1$), a direct consequence of its spectral bias. Furthermore, in later stages ($t_2$ and $t_3$), the unified model incorrectly estimates the low-frequency yaw motion, resulting in RMSE values exceeding $0.3$ across the frontal and rear regions. In contrast, the proposed framework accurately captures the localized deformation at $t_1$ and successfully reproduces the subsequent yaw motion at $t_2$ and $t_3$ with RMSE values below $0.21$. This performance confirms that the decomposition improves both temporal robustness and spatial precision by isolating the rigid-body and structural deformation components. The oracle model, which provides the analytical upper bound for deformation prediction by fixing the ground-truth rigid motion, shows RMSE fields that are closely aligned with those of the proposed framework.
\begin{figure}[H]
\centering
\includegraphics[width=0.8\linewidth]{Figure/figure_13.pdf}
\caption{
Comparison of RMSE fields for a validation scenario within the training parameter range, defined by $v = 58.21\,\text{km/h}$, $\theta = 22.5^\circ$, $r_{\text{offset}} = 51.35\%$, and $d = 1.07\,\text{m}$. Columns correspond to three time instants: $t_1 = 0.12\,\text{s}$ (initial impact), $t_2 = 0.20\,\text{s}$ (softening), and $t_3 = 0.36\,\text{s}$ (late impact). Rows show ground truth, unified (coupled MLP), oracle, and proposed models. The bottom panel presents the temporal RMSE evolution across all nodes, where the dashed line denotes the onset of impact.
}
\label{fig:interpolation_field}
\end{figure}

\begin{figure}[H]
\centering
\includegraphics[width=0.9\linewidth]{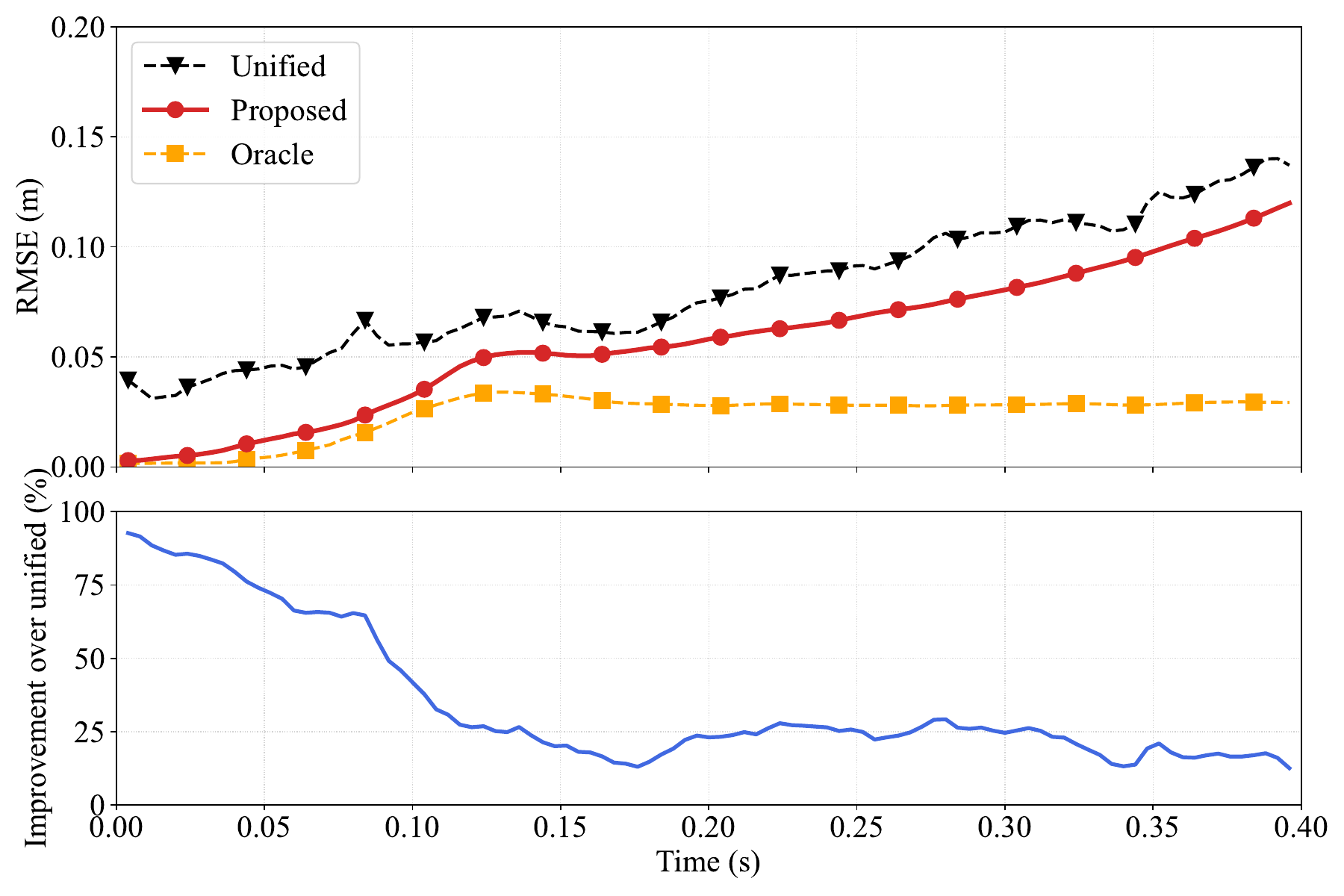}
\caption{
Overall interpolation performance across all validation scenarios. The upper panel reports the mean RMSE curves of the unified, oracle, and proposed models. The lower panel shows the relative improvement of the proposed framework over the unified model, expressed as the percentage RMSE reduction at each time step. The proposed model maintains consistently lower RMSE and remains comparable to the oracle throughout the sequence.
}
\label{fig:interpolation_overall_RMSE}
\end{figure}

Figure~\ref{fig:interpolation_overall_RMSE} summarizes the interpolation performance across all validation scenarios. The proposed model consistently maintains lower RMSE than the unified model over the entire sequence. The proposed framework closely approximates the accuracy of the oracle model throughout the validation set, demonstrating high efficacy even without access to ground-truth rigid motion. In addition to predictive accuracy, we evaluated computational efficiency. While the ground-truth FEM simulation for this scenario required $1.21\,\text{h}$, the proposed model predicts the full spatio-temporal dynamics in $247.06\,\text{ms}$. This performance corresponds to a speed-up factor of approximately $1.76 \times 10^4$, confirming the utility of the framework for rapid design iterations.

\subsection{Generalization to extrapolated scenarios}
\label{subsec:extrapolation}
We evaluated model performance under collision parameters outside the training ranges. We selected collision angle and velocity as extrapolation dimensions because they impose different physical challenges: changes in impact orientation versus changes in impact energy. The training data covered angles between $0^\circ$ and $45^\circ$ and velocities between $40$ and $80\,\text{km/h}$. Extrapolation was examined using unseen angles ($46^\circ$--$53^\circ$) and a high-velocity scenario ($97.0\,\text{km/h}$). Other parameters were fixed as $r_{\text{offset}} = 60.5\%$ and $d = 0.28\,\text{m}$.

\begin{figure}[H]
    \centering
    \includegraphics[width=0.8\linewidth]{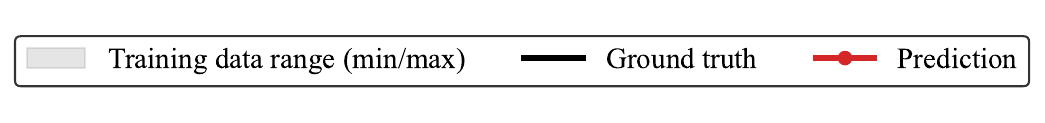}
    \begin{subfigure}[b]{0.48\linewidth}
        \includegraphics[width=\linewidth]{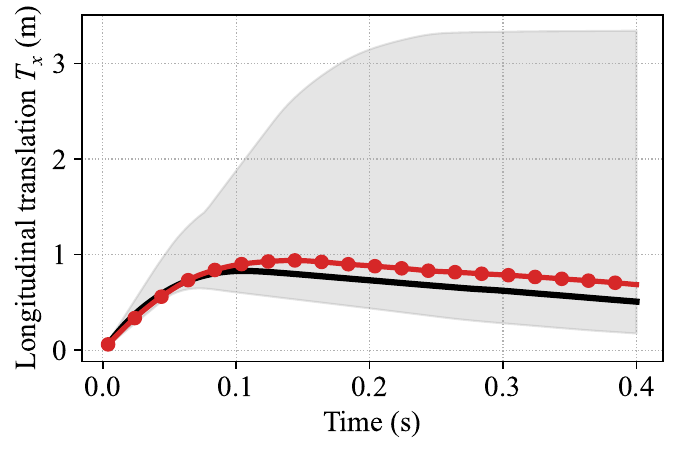}
        \caption{Angular extrapolation: $T_x$}
        \label{fig:angular_tx}
    \end{subfigure}
    \hfill
    \begin{subfigure}[b]{0.48\linewidth}
        \includegraphics[width=\linewidth]{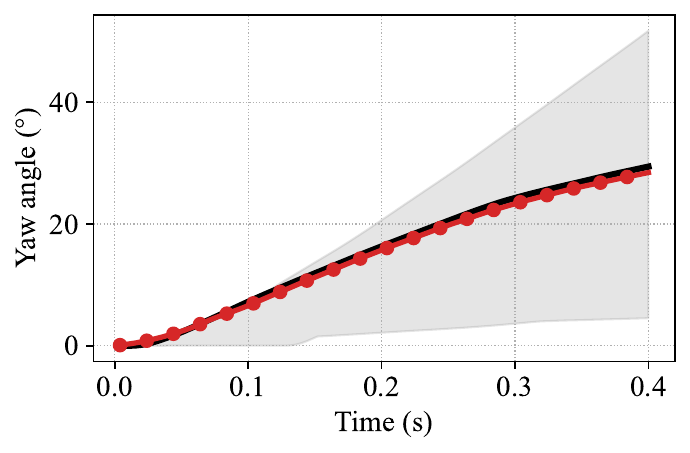}
        \caption{Angular extrapolation: yaw angle}
        \label{fig:angular_yaw}
    \end{subfigure}    

    \begin{subfigure}[b]{0.48\linewidth}
        \includegraphics[width=\linewidth]{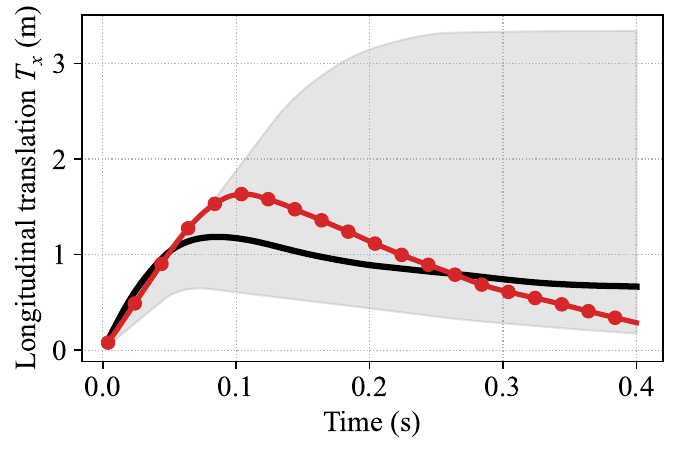}
        \caption{Velocity extrapolation: $T_x$}
        \label{fig:velocity_tx}
    \end{subfigure}
    \hfill
    \begin{subfigure}[b]{0.48\linewidth}
        \includegraphics[width=\linewidth]{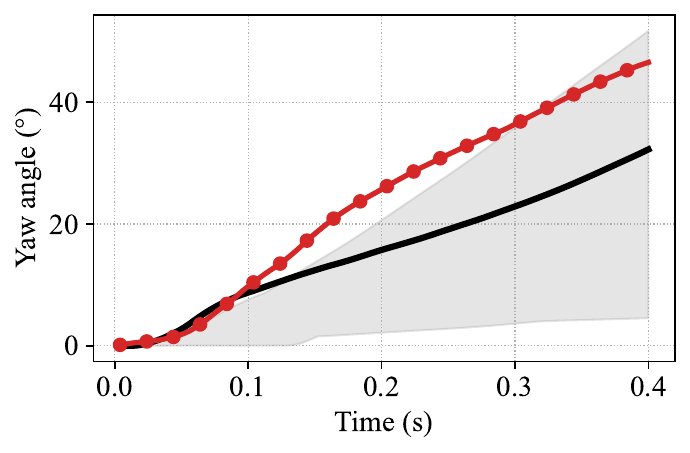}
        \caption{Velocity extrapolation: yaw angle}
        \label{fig:velocity_yaw}
    \end{subfigure}
    \caption{
        Analysis of extrapolation performance for \textit{RigidNet}. The model successfully generalizes under angular changes (a, b) but fails under high-velocity extrapolation (c, d). Notably in (d), the prediction matches the linear pre-crash phase but diverges during the non-linear impact phase due to the unseen energy regime.
    }
    \label{fig:extrapolation_rigid_motion}
\end{figure}

Since the proposed framework operates sequentially, the extrapolation capability is bounded by the stability of the first stage, \textit{RigidNet}. Figure~\ref{fig:extrapolation_rigid_motion} compares the predicted rigid-body trajectories against the training data envelope. For angular extrapolation (Figures~\ref{fig:angular_tx}, \ref{fig:angular_yaw}), \textit{RigidNet} generalizes. This is attributed to the fact that angular changes primarily alter the impact orientation while maintaining the total kinetic energy level comparable to the training distribution. Consequently, the physics remain within a regime that the network can resolve. In contrast, velocity extrapolation presents a severe challenge due to the quadratic scaling of kinetic energy ($E \propto v^2$). As shown in Figures~\ref{fig:velocity_tx} and \ref{fig:velocity_yaw}, the model exhibits a distinct failure mode. Before impact, the trajectory is predicted correctly, as the vehicle motion follows simple linear kinematics ($x=vt$) even at higher speeds. However, the trajectory diverges immediately after impact. This indicates that the model, trained only on the $40$--$80\,\text{km/h}$ range, fails to extrapolate the non-linear energy dissipation mechanics governing the high-speed crash phase. Because the underlying physics of structural crushing shifts significantly with the increased energy, the predicted yaw trajectory totally departs from the valid physical envelope.

\begin{figure}[H]
\centering
\includegraphics[width=0.9\linewidth]{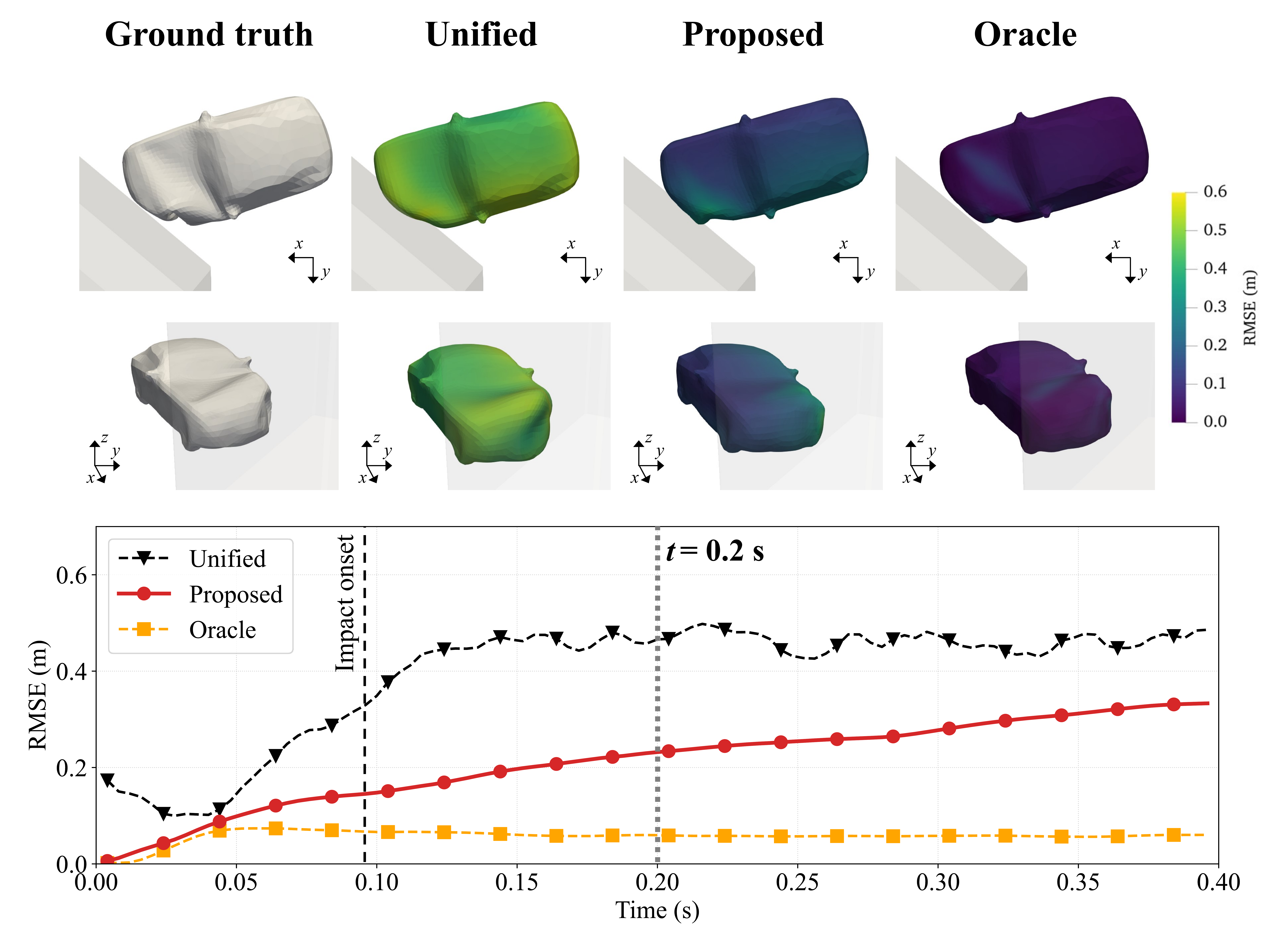}
\caption{
Total displacement RMSE fields for the angular extrapolation scenario ($\theta = 50.0^\circ$). The proposed framework maintains low error, demonstrating robustness to orientation variations.
}
\label{fig:extrapolation_angle_field}
\end{figure}

\begin{figure}[H]
\centering
\includegraphics[width=0.9\linewidth]{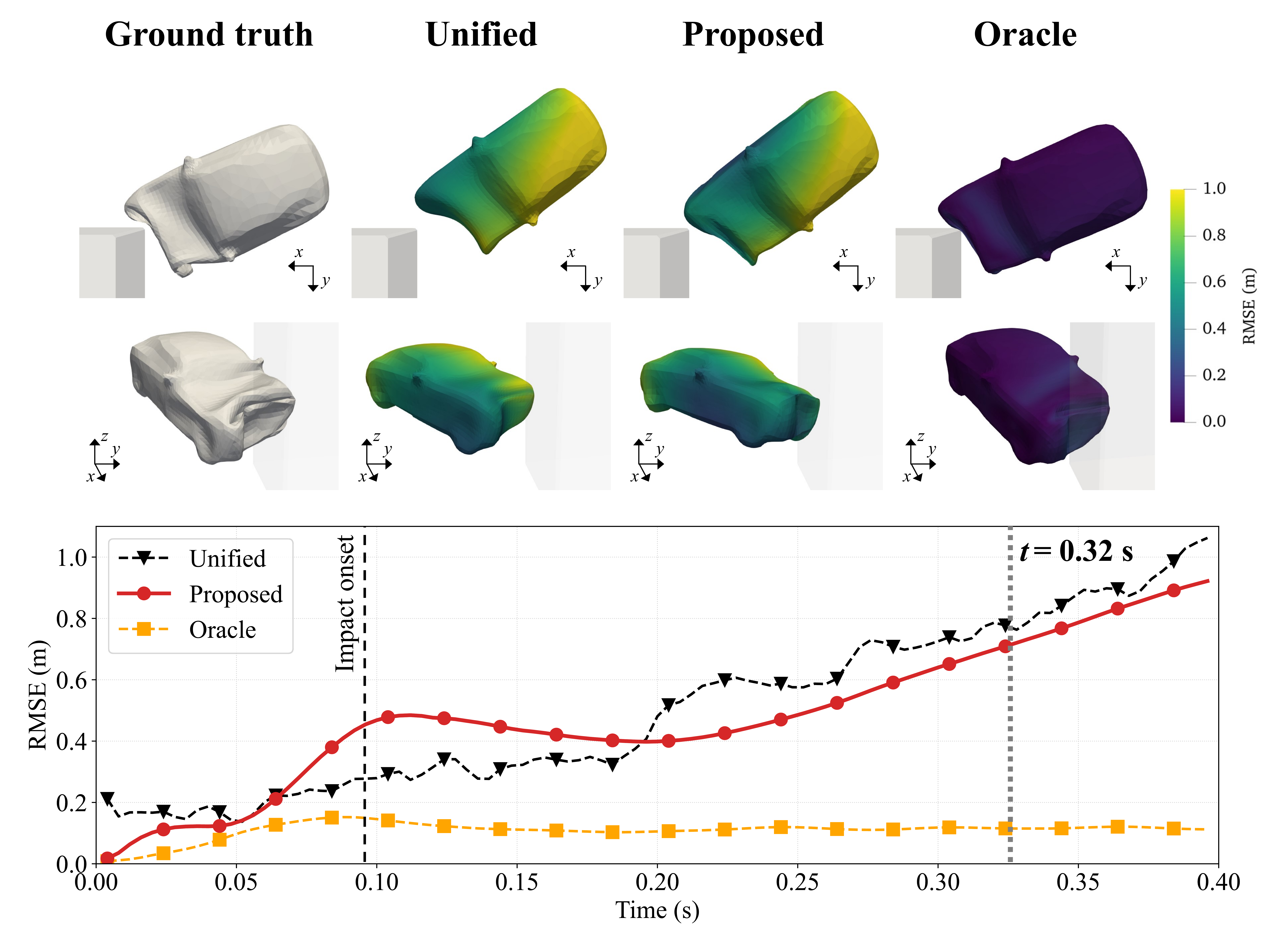}
\caption{
Total displacement RMSE fields for the velocity extrapolation scenario ($v = 97.0\,\text{km/h}$). The error amplification is a direct consequence of the \textit{RigidNet} failure to capture the high-energy non-linear dynamics.
}
\label{fig:extrapolation_velocity_field}
\end{figure}

The final total displacement accuracy directly reflects this sequential dependency. In the angular extrapolation scenario (Figure~\ref{fig:extrapolation_angle_field}), the stable rigid-body prediction translates to robust performance. The framework achieves a mean total displacement RMSE of $0.2072 \pm 0.0207$, representing a $46.6\%$ reduction compared to the unified model ($0.3878 \pm 0.0194$) and approaching the oracle performance. Conversely, in the velocity extrapolation scenario (Figure~\ref{fig:extrapolation_velocity_field}), the error accumulates rapidly during the post-impact phase, reaching approximately $1.0$. This error amplification is the direct consequence of the incorrect kinematic reference provided by \textit{RigidNet}. This contrast confirms that while the decomposition framework ensures accurate modeling within the learned energy regime, its extrapolation limit is bounded by \textit{RigidNet}'s capacity to resolve unseen non-linear dynamics. Future work to address this limitation could involve expanding the training domain to include high-energy regimes or incorporating energy-conservation constraints directly into the \textit{RigidNet} loss function.

\section{Physical validation of rigid-deformation decomposition}
\label{sec:verification}
This section validates the physical consistency of the proposed framework. We assess whether the separated components, rigid-body motion and structural deformation, adhere to the governing dynamics of vehicle collisions. The analysis compares the performance of the three training strategies defined in Section~\ref{subsubsec:ablation_training_strategies} to determine their impact on component isolation. Section~\ref{subsec:effect_training_strategies} quantifies the accuracy of the decomposed fields, demonstrating how the training strategy influences the decoupling of the learned components. Section~\ref{subsec:physical_consistency} verifies the consistency of the predictions based on three physical criteria: directional consistency, spatial localization, and temporal energy evolution.

\subsection{Impact of training strategy on component decoupling}
\label{subsec:effect_training_strategies}
We characterize the impact of the three training strategies: (D) baseline (end-to-end), (E) joint fine-tuning, and (F) proposed (frozen-anchor), on the decoupling of global rigid-body motion and residual deformation.

% \subsubsection{Accuracy of rigid-body motion}
\subsubsection{Stability of rigid-body motion against deformation interference}
\label{subsubsec:rigid_body_analysis}
\begin{figure}[H]
\centering
\includegraphics[width=1.0\linewidth]{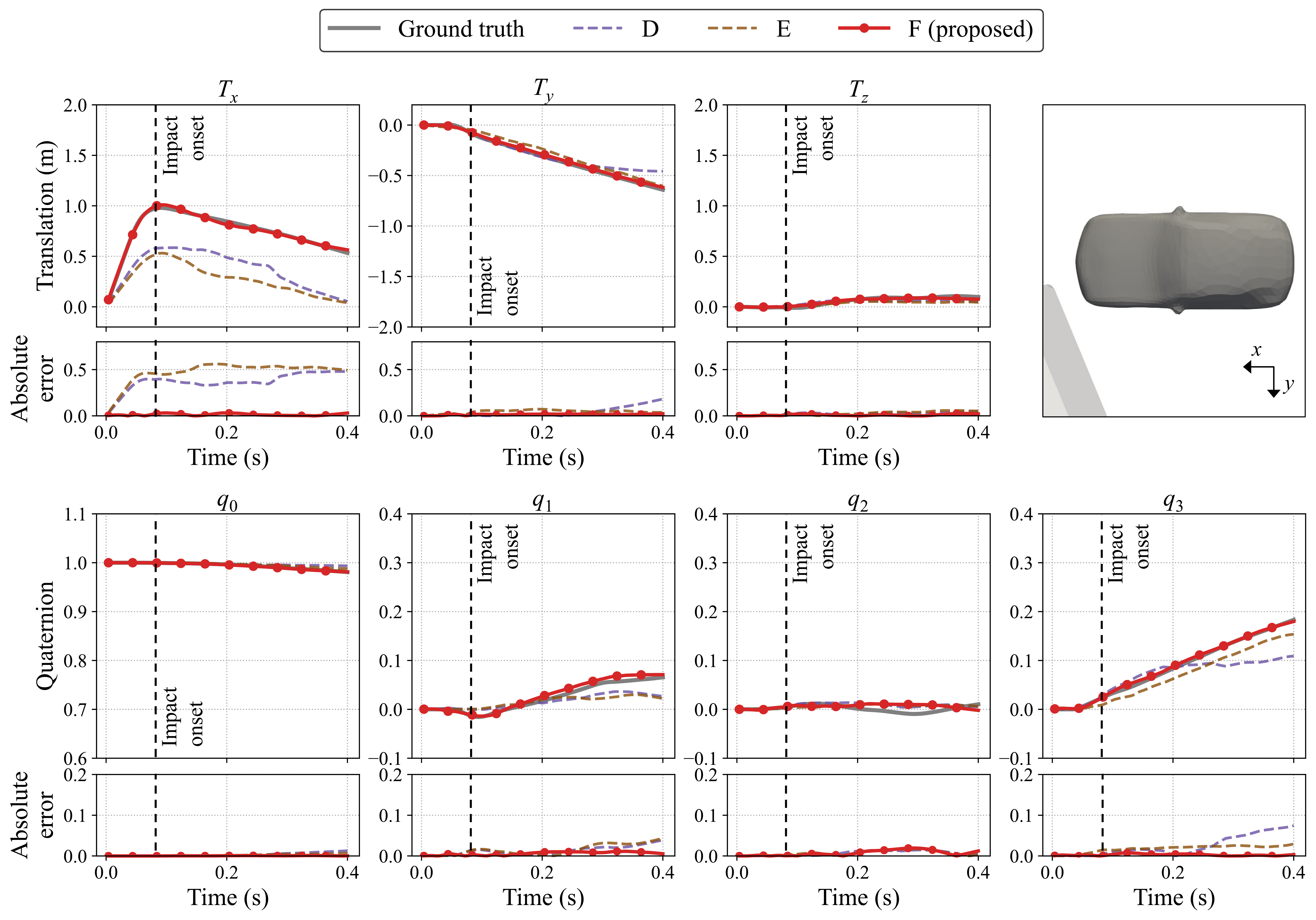}
\caption{
Time histories of translational ($T_x$, $T_y$, $T_z$) and rotational ($q_0$--$q_3$) components predicted by \textit{RigidNet} under the three training strategies. The frozen-anchor (F) strategy maintains bounded translation and quaternion trajectories, whereas end-to-end (D) and joint fine-tuning (E) strategies exhibit trajectory divergence due to deformation interference.
}
\label{fig:rigid_inference_time}
\end{figure}

The stability and accuracy of rigid-body motion are essential for the success of residual learning, as any error in the global prediction directly corrupts the moving reference frame for the subsequent deformation task. Therefore, we first evaluate the rigid-body motion predicted by \textit{RigidNet} for a representative validation scenario ($v = 58.21\,\text{km/h}$, $\theta = 22.5^\circ$, $r_{\text{offset}} = 51.35\%$, $d = 1.07\,\text{m}$). Figure~\ref{fig:rigid_inference_time} presents the time histories of the translational ($T_x, T_y, T_z$) and rotational ($q_0$--$q_3$) components. Longitudinal translation $T_x$, which corresponds to the primary axis of momentum change, exhibits divergence in strategies (D, E). These models show a divergence from the ground-truth trajectory after the initial impact, resulting in accumulated error. Similarly, the rotational components $q_1$ and $q_3$, associated with the vehicle's yaw motion during the late impact phase, show increasing deviations in strategies (D, E).

These deviations confirm a failure of joint training to isolate global dynamics. Specifically, the interference in (D, E) occurs because the coupled optimization encourages \textit{RigidNet} to overfit to local structural distortions rather than tracking the global center of mass. This phenomenon represents a leakage of high-frequency deformation dynamics into the low-frequency rigid-body prediction. Consequently, the predicted trajectory becomes physically inconsistent, as the network attempts to minimize total loss by non-physically shifting the global frame. In contrast, strategy (F) maintains a stable evolution for both translation and the quaternion components, adhering closely to the ground truth. This stability confirms that the frozen-anchor strategy acts as a spectral barrier, preventing deformation interference and ensuring that \textit{RigidNet} learns the global dynamics independently.

\begin{figure}[H]
\centering
\includegraphics[width=1.0\linewidth]{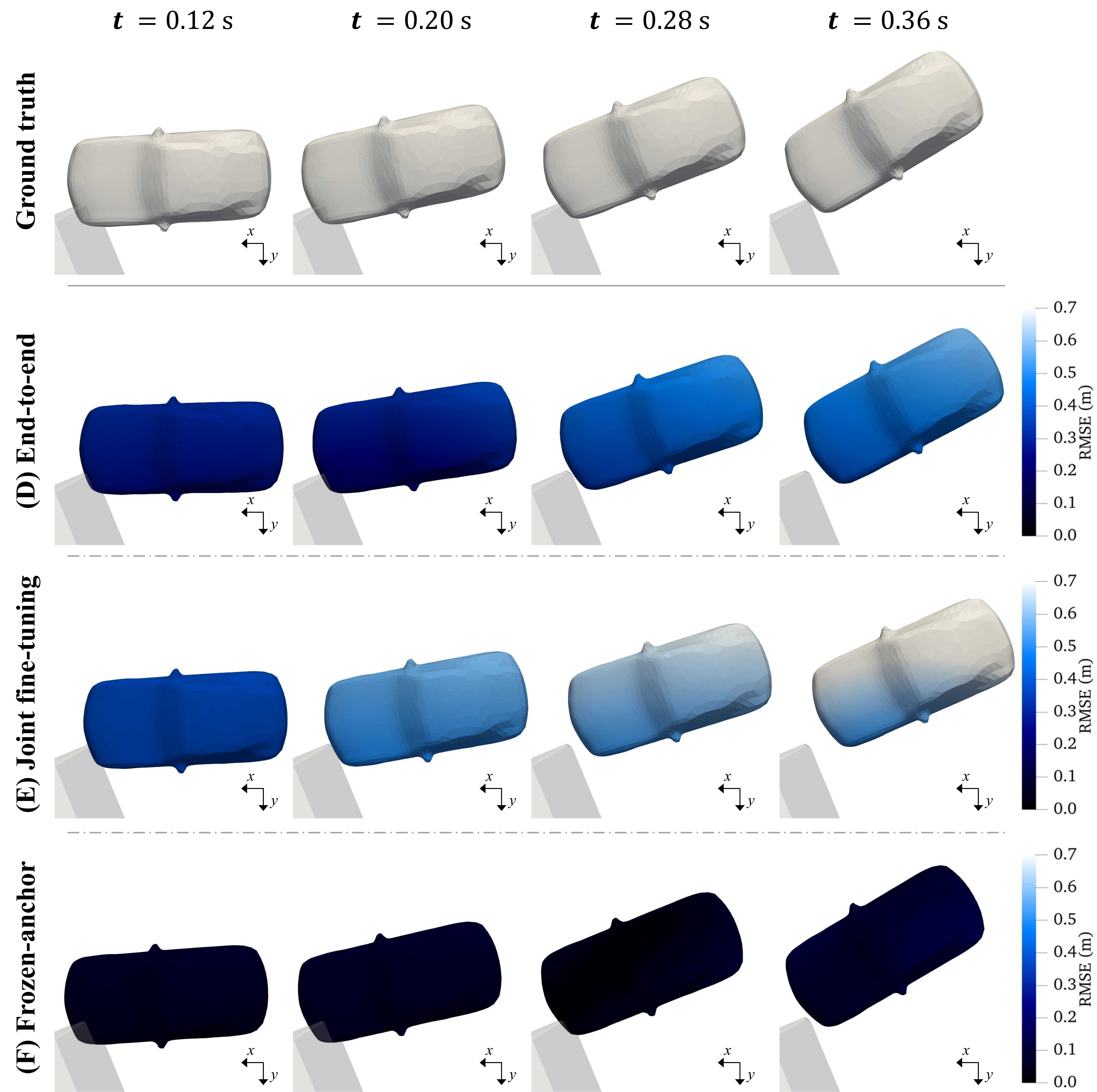}
\caption{
Spatial RMSE fields of \textit{RigidNet} under the three training strategies. Strategies (D, E) show concentrated errors in the rear region due to component coupling, while the frozen-anchor (F) strategy maintains low and uniform RMSE over time.
}
\label{fig:rigid_rmse_field}
\end{figure}

Figure~\ref{fig:rigid_rmse_field} presents the spatial RMSE field of \textit{RigidNet} at four time instances ($t = 0.12$, $0.20$, $0.28$, and $0.36\,\text{s}$). As the impact progresses, strategies (D, E) exhibit increasing errors concentrated in the rear region. The RMSE reaches approximately $0.5\,\text{m}$ for (D) and exceeds $0.7\,\text{m}$ for (E). This error concentration implies that the global coordinate system itself has deviated from the true trajectory. Instead of defining a proper kinematic anchor, the erroneous rigid motion forces the subtraction-based residual deformation to correct for global trajectory errors. This violates the principle of residual learning. In contrast, strategy (F) maintains an RMSE below $0.1\,\text{m}$ across all time steps, showing a uniform and stable distribution. This confirms that independent training ensures the physical reference remains purely kinematic, uncorrupted by local deformation.

\begin{figure}[H]
    \centering
    \includegraphics[width=0.35\linewidth]{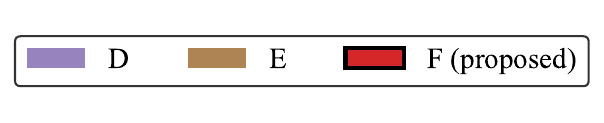}\\[1ex]
    \begin{subfigure}[b]{0.4\linewidth}
        \includegraphics[width=\linewidth]{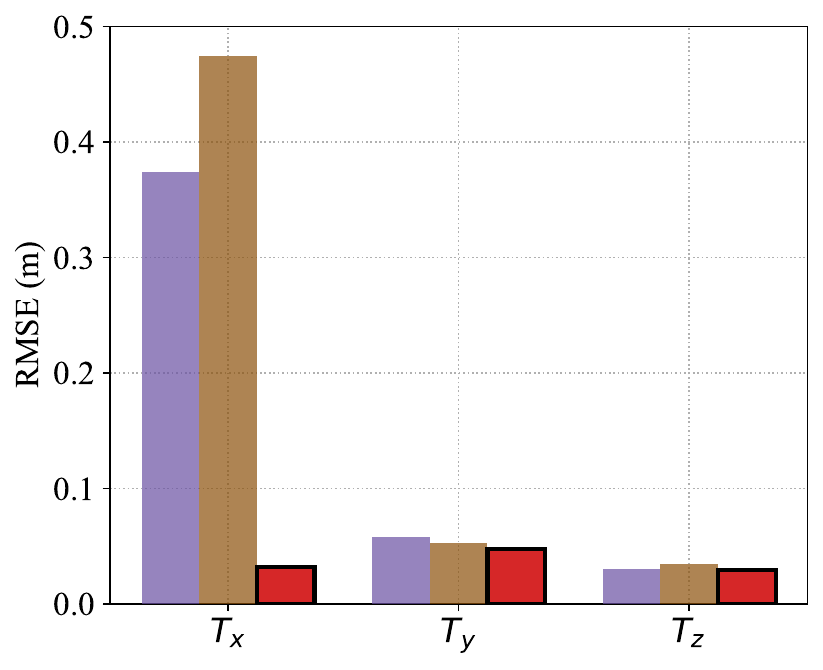}
        \caption{Translation}
        \label{fig:5_1_translation}
    \end{subfigure}
    \begin{subfigure}[b]{0.4\linewidth}
        \includegraphics[width=\linewidth]{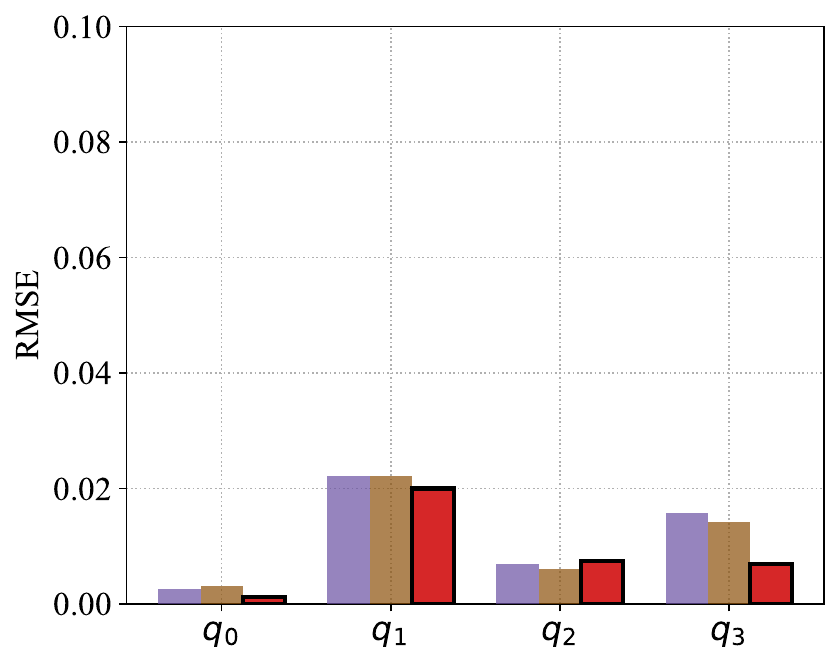}
        \caption{Rotation (quaternion)}
        \label{fig:5_1_quaternion}
    \end{subfigure}
    \caption{
    RMSE of translation and rotation, averaged over validation parameters and time steps. The frozen-anchor (F) strategy yields the lowest error in both components, indicating improved stability in global rigid-body motion.
    }
    \label{fig:summary_bar_plot}
\end{figure}

Figure~\ref{fig:summary_bar_plot} compares the mean RMSE values of translational and rotational components, averaged over all validation scenarios and time steps. Figure~\ref{fig:5_1_translation} presents the translation error, while Figure~\ref{fig:5_1_quaternion} shows the rotation error. Strategies (D, E) yield the largest error in the longitudinal direction ($T_x$). Since $T_x$ represents the primary axis of momentum transfer, this significant error indicates that coupled training compromises the tracking of global collision dynamics. Strategy (F) achieves the lowest RMSE in both translation and rotation, demonstrating that explicitly isolating the rigid-body optimization is essential for physically reliable long-term prediction.

\subsubsection{Robustness of local deformation against rigid-body trajectory errors}
\label{subsubsec:deformation_analysis}
Evaluating residual deformation is essential to determine whether the network captures physical distortion or merely compensates for global trajectory errors. We analyze the fields predicted by \textit{DeformationNet} using the same validation scenario described in Section~\ref{subsubsec:rigid_body_analysis}. Figure~\ref{fig:deform_verification_field} presents the spatial fields of predicted deformation magnitude (top) and RMSE (bottom) at $t = 0.20\,\text{s}$ (softening phase). Strategies (D, E) exhibit high RMSE values ($0.4$--$0.6\,\text{m}$) distributed across the entire vehicle domain. These errors stem from the unbounded kinematic divergence identified in the previous section. Since the rigid-body motion is overestimated, the residual field absorbs significant displacement components unrelated to structural distortion. Consequently, \textit{DeformationNet} is forced to compensate for these global errors, generating non-physical deformation artifacts in rigid zones, such as the rear body.

In contrast, strategy (F) maintains an RMSE below $0.05\,\text{m}$ and produces a deformation field localized within the frontal impact zone. By stabilizing the kinematic anchor, the strategy ensures that \textit{DeformationNet} exclusively learns the local structural response. The performance of strategy (F) aligns with that of the oracle model, which establishes the theoretical upper bound given ground-truth rigid motion. This alignment confirms that while concurrent training (D, E) degrades decoupling through interference, the frozen-anchor strategy (F) successfully isolates the components. The results indicate that the decomposition framework has reached its performance ceiling for deformation learning, with total accuracy now limited mainly by the precision of \textit{RigidNet}.
\begin{figure}[H]
\centering
\includegraphics[width=1.0\linewidth]{Figure/figure_21.pdf}
\caption{
Predicted deformation magnitude (top) and RMSE (bottom) fields at $t=0.20\,\text{s}$ for training strategies (D), (E), (F), and the oracle model. Strategies (D, E) exhibit non-physical deformation in rigid zones due to component coupling. The frozen-anchor strategy (F) eliminates these artifacts, yielding a localized response that closely matches the oracle's performance bound.
}
\label{fig:deform_verification_field}
\end{figure}

\subsection{Physical consistency of the proposed framework}
\label{subsec:physical_consistency}
While Section~\ref{subsec:effect_training_strategies} demonstrated high numerical accuracy, low error does not guarantee physical validity. This section assesses the physical consistency of the decomposition using three criteria: directional consistency (Section~\ref{subsubsec:directional_correlation}), spatial localization (Section~\ref{subsubsec:deformation_region}), and temporal energy evolution (Section~\ref{subsubsec:deformation_magnitude}). The analysis is conducted across eight validation scenarios to ensure robustness.

\subsubsection{Criterion 1: Directional consistency of decomposed components}
\label{subsubsec:directional_correlation}
In a frontal collision, the governing physics imposes a geometric constraint: as the vehicle penetrates the barrier, the global rigid-body motion follows the inertial trajectory forward (impact vector). Simultaneously, the structural nodes within the crush zone are pushed backward relative to the chassis. Consequently, the vectors representing these two components must be opposing. A physically valid decomposition implies that their dot product should be negative during the primary impact phase. Evaluating this relationship validates whether the model successfully decouples the inertial transport from the compressive structural response. To assess this alignment independent of scale, we quantify the directional consistency, \(\bar{C}\), using the cosine similarity between the predicted rigid-body vector \(\hat{\mathbf{r}}_{i}\) and deformation vector \(\hat{\mathbf{D}}_{i}\). The averaged metric is defined as:
\begin{equation}
\bar{C}(t) =
\mathrm{mean}\!\left(
\frac{
\hat{\mathbf{r}}_{i}(t) \cdot \hat{\mathbf{D}}_{i}(t)
}{
\left\|\hat{\mathbf{r}}_{i}(t)\right\|\,
\left\|\hat{\mathbf{D}}_{i}(t)\right\|
}
\right),
\end{equation}
where \(\bar{C}(t) < 0\) indicates counter-directional motion characteristic of physical compression, while \(\bar{C}(t) \ge 0\) implies uncorrelated or co-directional motion.

\begin{figure}[H]
    \centering
    \includegraphics[width=0.7\linewidth]{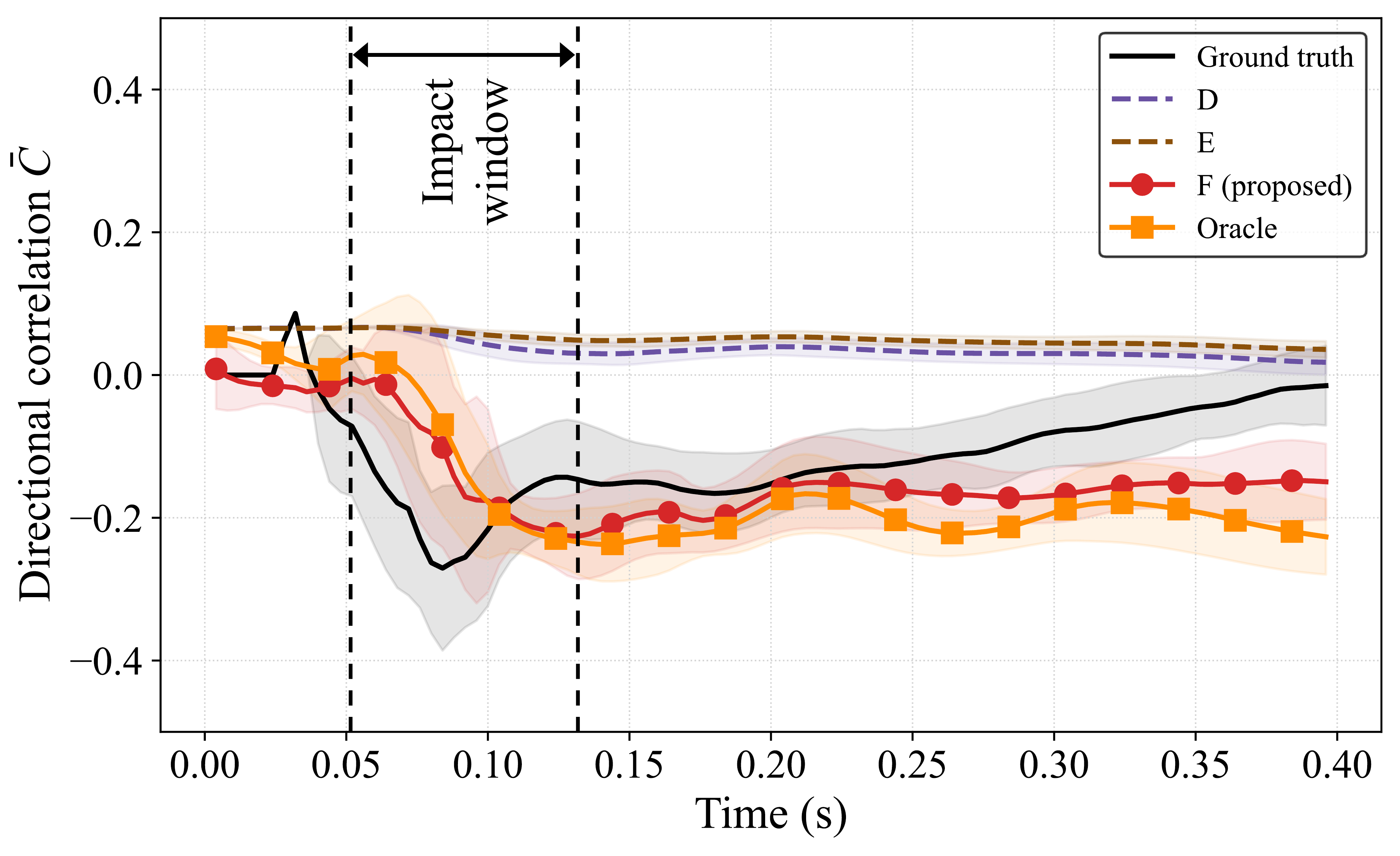}
    \caption{
    Temporal evolution of the directional correlation \(\bar{C}\). Solid curves and shaded regions represent the mean and standard deviation across eight validation scenarios, with the black curve denoting the ground truth. Vertical dashed lines mark the impact window. The frozen-anchor (F) model reproduces the negative correlation peak characteristic of the ground truth, whereas the end-to-end (D) and joint fine-tuning (E) models remain near zero due to component leakage.
    }
    \label{fig:correlation_overall}
\end{figure}

Figure~\ref{fig:correlation_overall} illustrates the temporal evolution of $\bar{C}$. The ground truth reveals a distinct physical signature: it drops rapidly during the early compression stage, reaching a peak negative correlation of approximately $-0.28$. This confirms that real-world deformation acts in opposition to rigid motion. The frozen-anchor (F) model reproduces this critical behavior, capturing approximately $92\%$ of the peak magnitude established by the oracle model ($0.22$ compared to $0.24$). In stark contrast, the end-to-end (D) and joint fine-tuning (E) strategies yield values hovering near zero throughout the entire duration. This lack of negative correlation provides empirical evidence of functional coupling. Since $\bar{C} \approx 0$ implies that the predicted deformation is directionally uncorrelated with the impact physics, it indicates that \textit{DeformationNet} in strategies (D, E) is not learning structural compression. Instead, it functions merely as a residual compensator, adjusting for \textit{RigidNet}'s positional errors in arbitrary directions. The proposed frozen-anchor strategy prevents this functional ambiguity, forcing \textit{DeformationNet} to strictly learn the counter-directional compressive dynamics consistent with the oracle.

\subsubsection{Criterion 2: Spatial consistency of deformation}
\label{subsubsec:deformation_region}
\begin{figure}[H]
    \centering
    \begin{subfigure}[t]{0.3\linewidth}
        \centering
        \includegraphics[width=\linewidth]{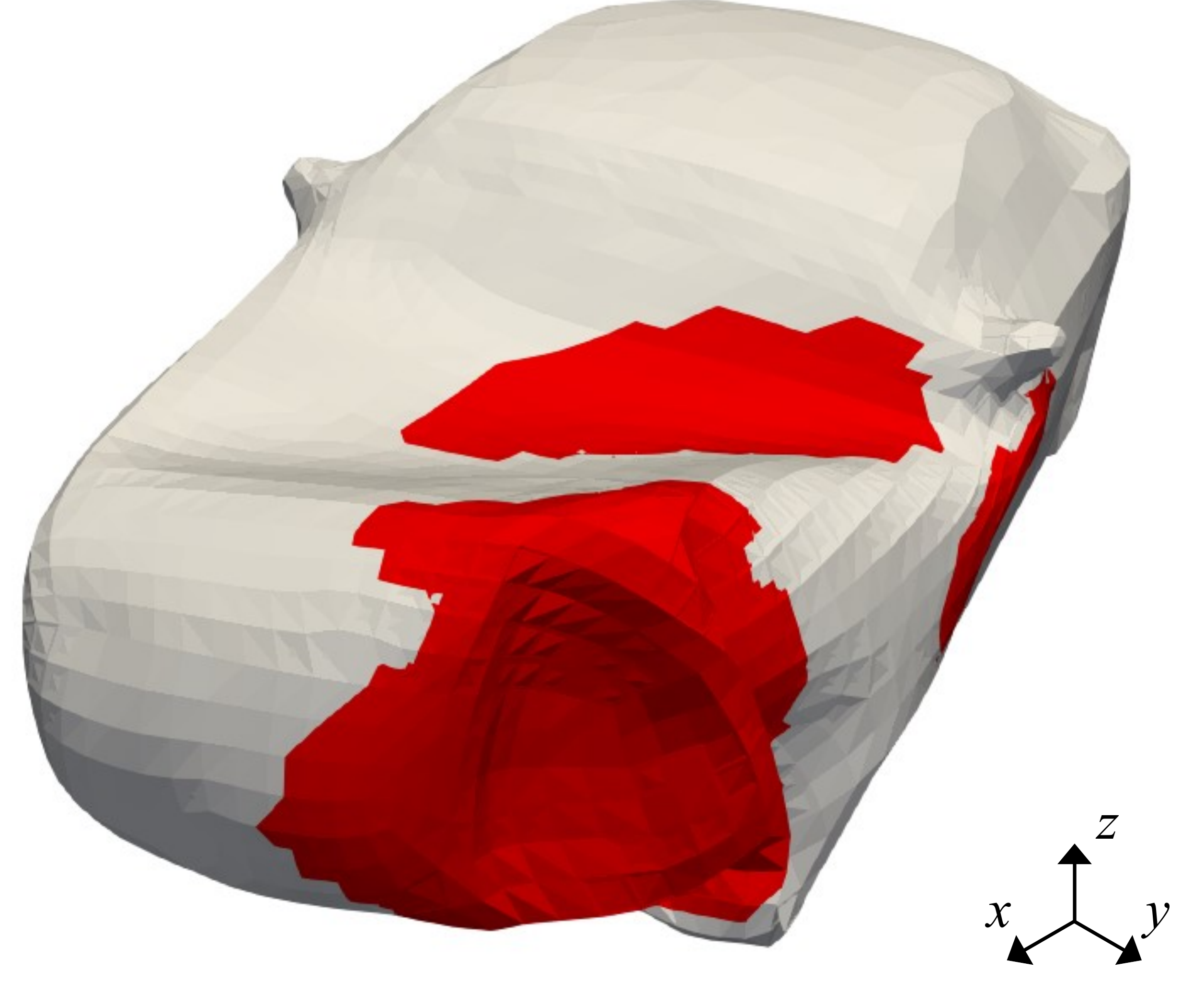}
        \caption{Ground-truth high-deformation region (top $15\%$)}
        \label{fig:deformation_region_a}
    \end{subfigure}
    \begin{subfigure}[t]{0.50\linewidth}
        \centering
        \includegraphics[width=\linewidth]{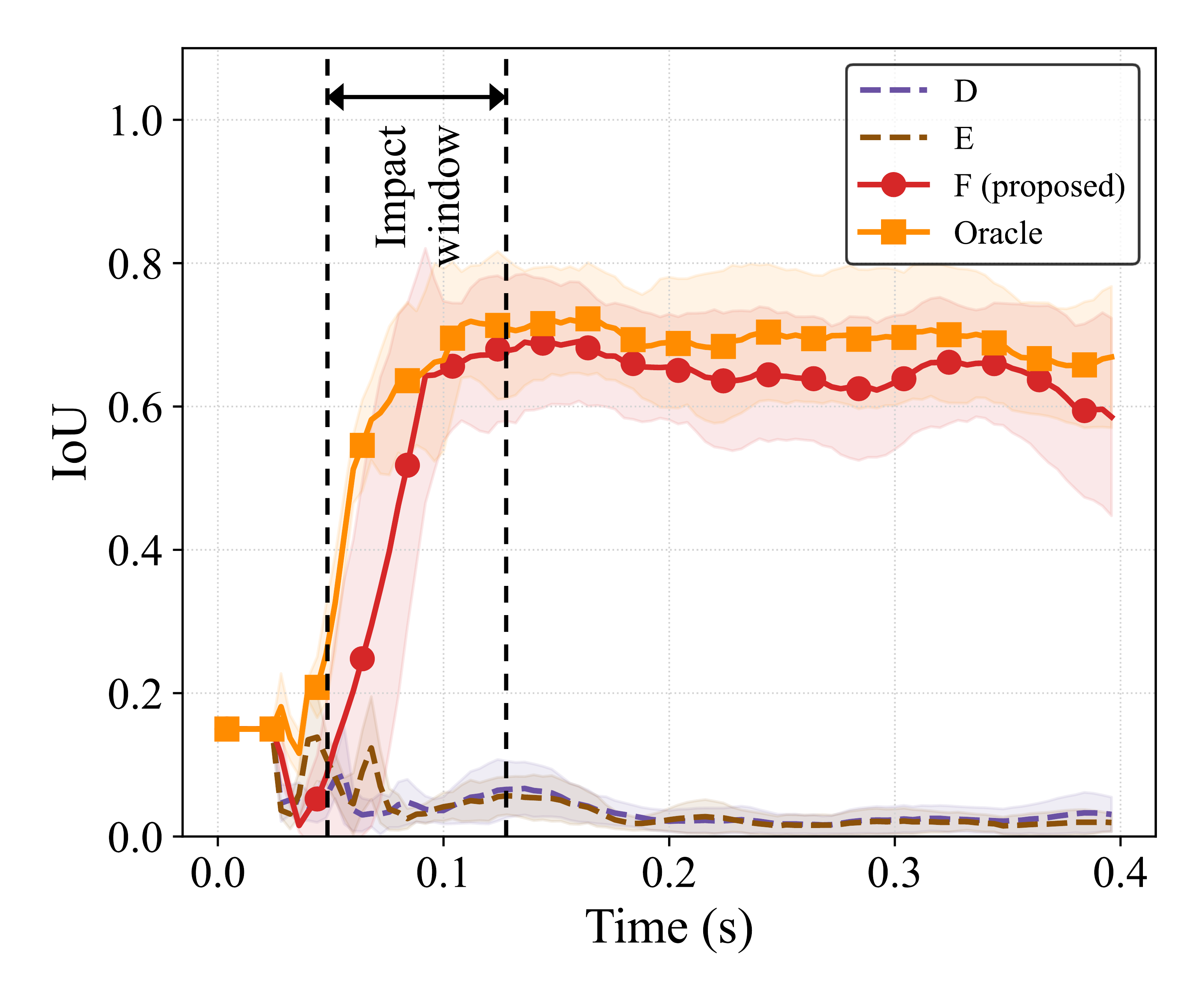}
        \caption{Temporal evolution of IoU for high-deformation region}
        \label{fig:deformation_region_b}
    \end{subfigure}
    \caption{
    Spatial correspondence of the high-deformation regions. (a) Visualization of the active crush zone (red), defined as the top $15\%$ deformed nodes. (b) Temporal evolution of IoU. Strategies (D) and (E) fail to localize the damage (near-zero IoU), while the frozen-anchor (F) model achieves high overlap with the ground truth, closely matching the oracle's performance.
    }
    \label{fig:deformation_region_scenario5}
\end{figure}

In vehicle collisions, kinetic energy dissipation is spatially concentrated within the crumple zone. A physically valid model must reproduce this localization; predicting deformation outside this zone implies a failure to capture the primary load path. To quantify the spatial consistency of the failure mechanism, we treat deformation prediction as a region detection task and utilize the Intersection over Union (IoU) metric. We define the active crush zone, $\Omega_{\mathrm{GT}}$, as the set of nodes within the top $15\%$ of deformation magnitude (visualized in Figure~\ref{fig:deformation_region_a}). The predicted set, $\Omega_{\mathrm{pred}}$, is derived similarly. The IoU is calculated as:
\begin{equation}
\mathrm{IoU}(t) =
\frac{|\Omega_{\mathrm{pred}}(t) \cap \Omega_{\mathrm{GT}}(t)|}
{|\Omega_{\mathrm{pred}}(t) \cup \Omega_{\mathrm{GT}}(t)|}.
\label{eq:iou}
\end{equation}
This metric measures whether the model focuses its capacity on the physically damaged region or scatters predictions elsewhere.

Figure~\ref{fig:deformation_region_b} compares the temporal evolution of localization performance. Strategies (D) and (E) fail, exhibiting IoU scores near zero throughout the impact. This confirms that their predicted deformation is spatially disjoint from the actual crush zone. In contrast, strategy (F) demonstrates localization, with the IoU rising to a peak of approximately $0.70$. This performance corresponds to $96\%$ of the accuracy achieved by the oracle model ($0.73$). This close alignment confirms that the frozen-anchor strategy suppresses spatial noise, enabling \textit{DeformationNet} to localize deformation dynamics to the physically correct structural members.

\subsubsection{Criterion 3: Temporal consistency of deformation}
\label{subsubsec:deformation_magnitude}
The final criterion assesses whether the predicted deformation follows the characteristic temporal progression of a collision event: rapid accumulation, transient relaxation, and post-impact stabilization. We quantify this using the normalized deformation magnitude, $\tilde{M}$, which serves as a proxy for the total deformation energy state. It is defined as the node-averaged squared displacement norm, normalized by its peak value:
\begin{equation}
    \tilde{M}(t) = \frac{1}{M_{\mathrm{peak}}} \left( \frac{1}{N} \sum_{i=1}^{N} \| \mathbf{D}_i(t) \|^{2} \right),
    \label{eq:mag_norm}
\end{equation}
where $N$ is the total number of nodes and $M_{\mathrm{peak}}$ is the maximum magnitude observed in the ground truth. Using this metric, we dissect the collision dynamics into three distinct phases based on the rate of energy change. First, deformation accumulation (Phase I), defined by $\frac{d\tilde{M}}{dt} > 0$, corresponds to the rapid increase in magnitude driven by kinetic energy transfer. Second, post-peak transient response (Phase II), defined by $\frac{d\tilde{M}}{dt} < 0$, represents structural relaxation as momentum dissipates. Finally, post-impact stabilization (Phase III) is identified when the temporal variation satisfies $\left|\frac{d\tilde{M}}{dt}\right| \le \epsilon$ (with $\epsilon = 0.05$) for at least $0.02\,\text{s}$, marking convergence to a steady residual state.

\begin{figure}[H]
    \centering
    \includegraphics[width=0.6\textwidth]{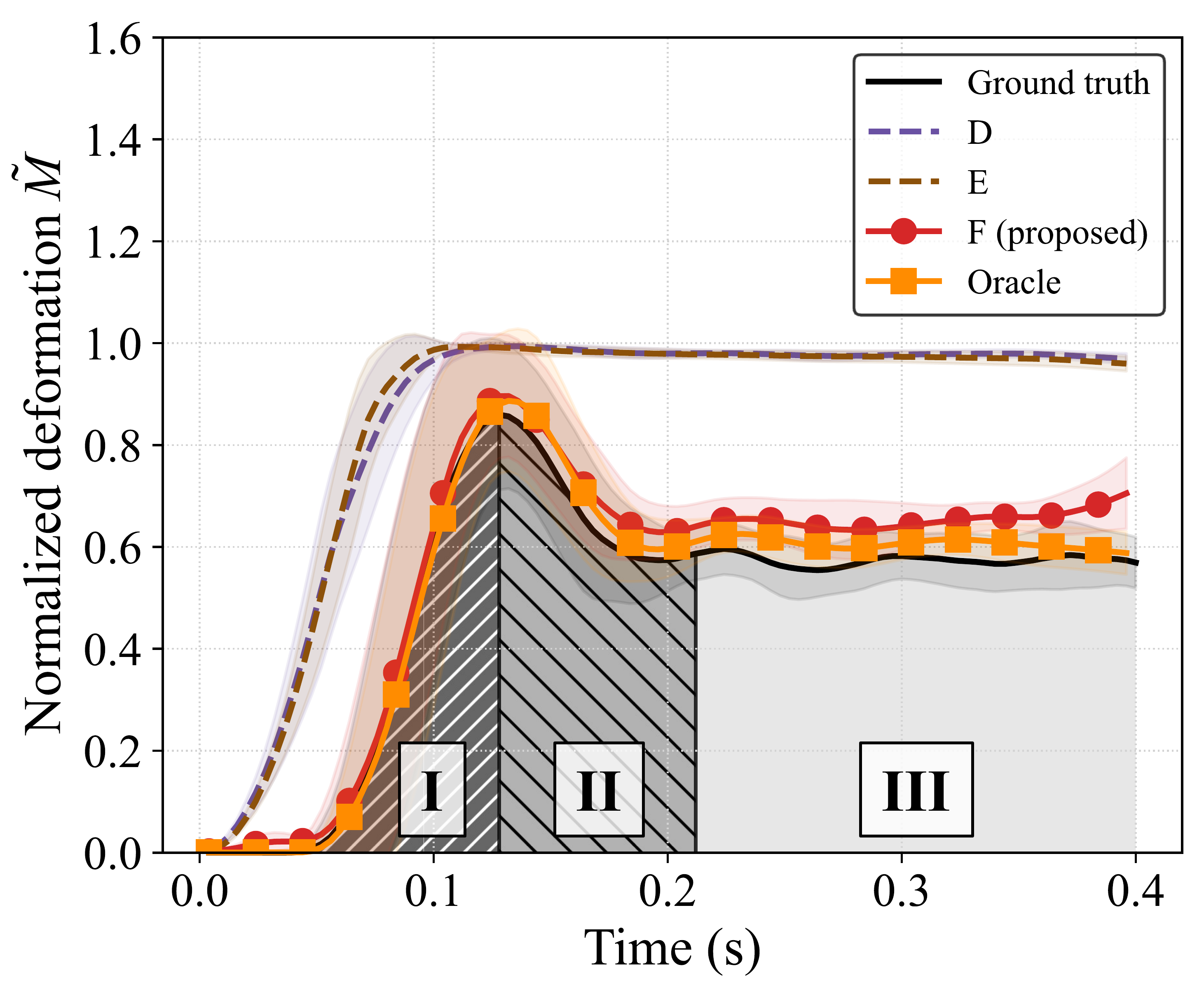}
    \caption{
    Temporal evolution of the normalized deformation magnitude $\tilde{M}$. The frozen-anchor strategy (F) and the oracle reproduce both the peak and the stabilization behavior observed in the ground truth, whereas strategies (D, E) retain elevated deformation and fail to stabilize; shaded regions denote standard deviation.
    }
    \label{fig:normalized_deform_mag}
\end{figure}

Figure~\ref{fig:normalized_deform_mag} illustrates the temporal evolution of $\tilde{M}$, where a progression of all three phases is shown. As quantified in Table~\ref{tab:phase_transition}, ground truth reaches peak accumulation at $0.128\,\text{s}$ and achieves stabilization by $0.212\,\text{s}$. Strategies (D, E) fail to reproduce this timeline: they exhibit peaks ($0.112\,\text{s}$ and $0.110\,\text{s}$) and fail to reach the stabilization phase (Phase III), as indicated by the lack of defined onset times in Table~\ref{tab:phase_transition}.  This failure is a direct consequence of the trajectory divergence identified in Section~\ref{subsubsec:rigid_body_analysis}. Because the rigid-body error accumulates over time, the subtraction-based deformation field inherits a continuous, non-physical deviation, preventing the residual energy from stabilizing even after the collision concludes. In contrast, the frozen-anchor strategy (F) and the oracle model exhibit temporal alignment with the ground truth. Both models capture the Phase I peak at $0.132\,\text{s}$ and enter Phase III with a negligible delay of $8\,\text{ms}$ ($0.220\,\text{s}$ compared to $0.212\,\text{s}$). This correspondence confirms that eliminating kinematic interference via the frozen-anchor strategy allows \textit{DeformationNet} to correctly learn the energy dissipation characteristics, accurately reproducing the peak response, decay, and final stabilization of the structural system.

\begin{table}[H]
\centering
\caption{Transition times of the normalized deformation magnitude for each model.}
\label{tab:phase_transition}
\begin{tabular}{l rr}
\hline
\makecell{\textbf{Model}} &
\makecell{\textbf{Phase I peak time (s)}} &
\makecell{\textbf{Phase III onset (s)}} \\
\hline
Ground truth     & 0.128 & 0.212 \\
D                & 0.112 & -- \\
E                & 0.110 & -- \\
\textbf{F (proposed)} & \textbf{0.132} & \textbf{0.220} \\
Oracle           & 0.132 & 0.220 \\
\hline
\end{tabular}
\end{table}

\section{Conclusion}
\label{sec:conclusion}
This study addressed the challenge of predicting vehicle collision dynamics characterized by coupled multi-scale responses. Single-network architectures fail to resolve low-frequency rigid-body motion and high-frequency structural deformation due to spectral bias. To overcome this limitation, we proposed a hierarchical framework based on rigid-deformation decomposition. Evaluations spanning interpolation, extrapolation, and physical consistency metrics were conducted to validate the efficacy of the proposed approach. The key findings and contributions are summarized as follows:
\begin{itemize}
    \item \textbf{Decoupling of rigid-body motion and local deformation:} We established a framework integrating temporal incremental prediction using quaternions with a frozen-anchor training strategy. While joint training strategies failed to isolate dynamics due to kinematic leakage, our approach ensured decoupling of global and local components. Specifically, the incremental scheme reduced rigid-body prediction error by $29.8\%$ compared to conventional direct prediction, and the frozen-anchor strategy eliminated trajectory divergence, securing a stable kinematic reference.

    \item \textbf{Optimization landscape smoothing and generalization:} We demonstrated that the decomposition architecture smooths the optimization landscape. Visualization revealed that while single-network models converge to sharp minima, the proposed framework settles into a flat basin. This property improves robustness, enabling the model to outperform baselines with a $17.2\%$ reduction in interpolation error and a $46.6\%$ reduction in extrapolation error.
    
    \item \textbf{Verification of physical consistency}: We verified the physical consistency of the learned dynamics by benchmarking against an oracle model, which defines the performance upper bound. The framework satisfied three criteria relative to this bound: (1) capturing $92\%$ of the peak counter-directional signature; (2) achieving $96\%$ localization accuracy (IoU of $0.70$) within the crush zone; and (3) matching the temporal energy evolution with a delay of $8\,\text{ms}$. These results confirm that the proposed decomposition utilizes the network capacity to capture multi-scale physical dynamics beyond data-driven trends.
\end{itemize}

Despite these promising results, this study identified a distinct limitation regarding generalization to unseen kinetic energy regimes. While the model robustly generalizes to geometric variations such as collision angles, it exhibits performance degradation in velocity extrapolation scenarios. This failure indicates that the current kinematic decomposition strategy, although effective for resolving spectral bias, does not inherently capture the non-linear dynamic relationship between impact velocity and total energy dissipation. Consequently, the model struggles to extrapolate to energy states that lie outside the training distribution.

To address this challenge, future work will transition the current decomposition architecture toward a physics-informed framework. Specifically, we aim to integrate governing physical constraints, such as conservation laws of mass, momentum, and energy, directly into the loss function. By enforcing these fundamental dynamic principles, we aim to constrain the solution space in extrapolated domains, thereby ensuring physical consistency across varying energy regimes and advancing the reliability of deep learning-driven simulation in engineering design.

\section*{CRediT authorship contribution statement}
Sanghyuk Kim: Writing – review and editing, Writing – original draft, Visualization, Validation, Software, Methodology, Data curation, Conceptualization.
Minsik Seo: Conceptualization, Writing – review and editing.
Sunwoong Yang: Writing – review and editing, Investigation, Methodology, Conceptualization, Supervision.
Namwoo Kang: Writing – review and editing, Project administration, Funding acquisition, Supervision.

\section*{Declaration of Competing Interest}
The authors declare that they have no known competing financial interests or personal relationships that could have appeared to influence the work reported in this paper.

\section*{Acknowledgments}
This work was supported by the Ministry of Science and ICT of Korea (Grant Nos. 2022-0-00969, 2022-0-00986, and GTL24031-000) and the Ministry of Trade, Industry \& Energy (RS-2024-00410810 and RS-2025-02317327). Additional support was provided by the InnoCORE program of the Ministry of Science and ICT (N10250154).

\section*{Data Availability Statement}
The data that support the findings of this study are available from the corresponding author upon reasonable request.

\bibliography{reference}

\appendix

\section{Finite element simulation setup}
\label{appendix:fem_setup}
\renewcommand{\thefigure}{A.\arabic{figure}}
\renewcommand{\thetable}{A.\arabic{table}}
\setcounter{figure}{0}
\setcounter{table}{0}

Vehicle collision simulations were performed using Ansys Explicit Dynamics (AutoDyn solver). The aluminum alloy of the vehicle body was modeled as an elasto-plastic material with bilinear isotropic hardening. The concrete barrier was defined using the CONC-35MPA material model, characterized by properties such as a 35\,MPa compressive strength, as shown in Table~\ref{tab:grouped_material}. The simulation setup followed crashworthiness analysis protocols with established material models and contact algorithms.
\begin{table}[H]
\centering
\caption{Finite element simulation parameters for vehicle collision analysis using Ansys Explicit Dynamics (AutoDyn solver).}
\label{tab:grouped_material}
\begin{tabular}{l l r}
\hline
\makecell{\textbf{Category}} &
\makecell{\textbf{Description}} &
\makecell{\textbf{Value}} \\
\hline
\multirow{3}{*}{Geometry}
  & Shell element thickness               & 3 mm \\
  & Base mesh size                        & 0.3 m \\
  & Number of nodes/elements              & 2{,}567 / 2{,}606 \\
\hline
\multirow{12}{*}{Material}
  & \textbf{\textit{Vehicle: Aluminum Alloy NL}} & \\
  & Density                                   & 2{,}770 kg/m$^3$ \\
  & Young's modulus                           & 71 GPa \\
  & Poisson's ratio                           & 0.33 \\
  & Yield strength                            & 280 MPa \\
  & Tangent modulus                           & 500 MPa \\
  \cline{2-3}
  & \textbf{\textit{Barrier: CONC-35MPA}}     & \\
  & Density                                   & 2{,}314 kg/m$^3$ \\
  & Compressive strength                      & 35 MPa \\
  & Shear modulus                             & 16.7 GPa \\
  & Solid density                             & 2{,}750 kg/m$^3$ \\
\hline
\multirow{3}{*}{Analysis}
  & Minimum time step                         & $10^{-9}$ s \\
  & Time step safety factor                   & 0.9 \\
  & Friction coefficient                       & 0.8 (static), 0.7 (kinetic) \\
\hline
\end{tabular}
\end{table}

\section{Sensitivity analysis of training data size}
\label{appendix:data_sensitivity}
\renewcommand{\thefigure}{B.\arabic{figure}}
\renewcommand{\thetable}{B.\arabic{table}}
\setcounter{figure}{0}
\setcounter{table}{0}

We conducted a sensitivity analysis to examine the effect of training data size on model stability and accuracy. All experiments used the unified model (coupled MLP) described in Section~\ref{subsec:comparative_models} without spatio-temporal subsampling. We generated datasets via LHS based on the four collision parameters defined in Section~\ref{subsec:dataset_generation}: $v$, $\theta$, $r_{\text{offset}}$, and $d$. We compared performance across five training sample sizes, denoted as $N_{\text{DP}} \in \{10, 20, 30, 40, 50\}$.

Table~\ref{tab:appendix_dp_stats} summarizes the validation loss and average computation time per epoch for each configuration. The validation loss reached its minimum value of $0.00209$ at $N_{\text{DP}}=30$. However, increasing the sample size from 20 to 30 yielded only an accuracy improvement of approximately $7.1\%$ (from 0.00225 to 0.00209), while increasing the computational cost by $74.7\%$ (from $60.48\,\text{s}$ to $105.64\,\text{s}$). Furthermore, the model exhibited instability at $N_{\text{DP}} \ge 40$, where the validation loss increased. Considering the trade-off between predictive accuracy and computational efficiency, we selected 20 data points as a suitable balance for the subsequent experiments.
\begin{table}[H]
\centering
\caption{Validation loss and computation time per epoch for different training data sizes.}
\label{tab:appendix_dp_stats}
\begin{tabular}{r r r}
\hline
\makecell{\textbf{$N_{\text{DP}}$}} &
\makecell{\textbf{Best validation loss}} &
\makecell{\textbf{Time per epoch (s)}} \\
\hline
10 & 0.00296 & 28.71 \\
\textbf{20} & \textbf{0.00225} & \textbf{60.48} \\
30 & 0.00209 & 105.64 \\
40 & 0.00264 & 124.35 \\
50 & 0.00225 & 173.68 \\
\hline
\end{tabular}
\end{table}

\begin{figure}[H]
    \centering
    \includegraphics[width=0.85\linewidth]{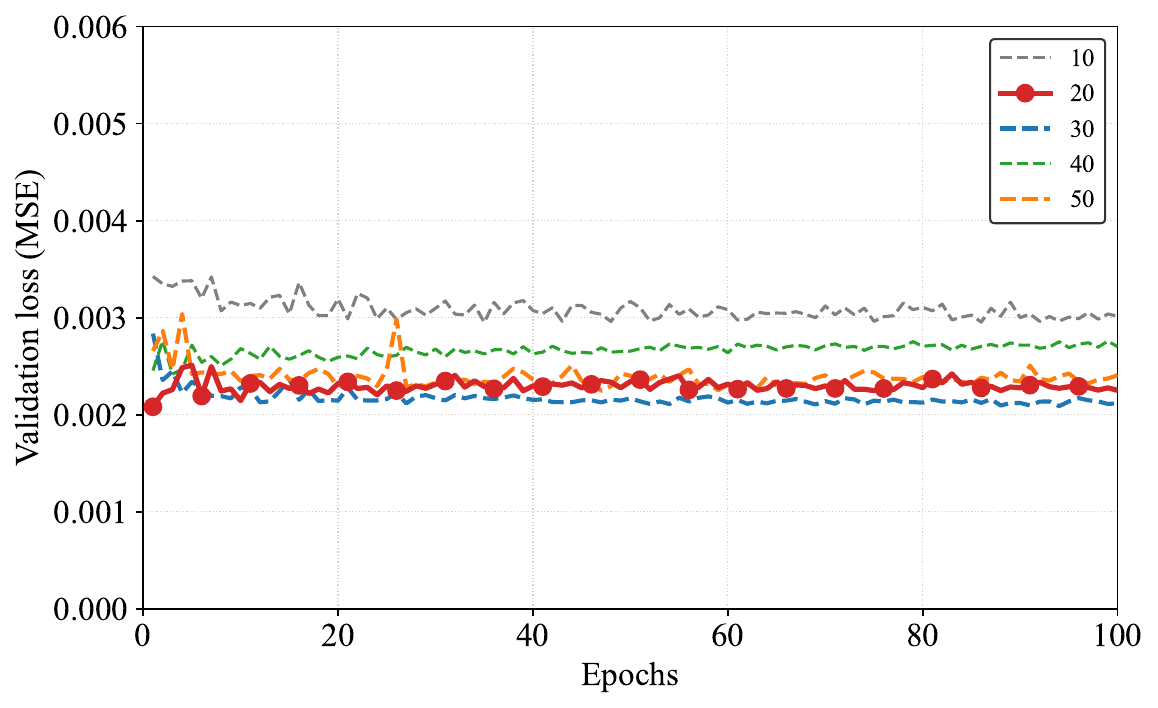}
    \caption{
    Validation loss curves for different training data sizes ($N_{\text{DP}} = 10, 20, 30, 40, 50$). All experiments were performed using the unified model (coupled MLP) without spatio-temporal subsampling. Each dataset was generated via LHS within the parameter ranges defined in Section~\ref{subsec:dataset_generation}. The performance improved up to $N_{\text{DP}}=20$ and plateaued thereafter, indicating that 20 data points were sufficient for stable learning.
    }
    \label{fig:appendix_dp_sensitivity}
\end{figure}

Figure~\ref{fig:appendix_dp_sensitivity} shows the validation loss curves up to 100 epochs for different values of $N_{\text{DP}}$. The loss decreased as $N_{\text{DP}}$ increased from 10 to 20. While $N_{\text{DP}}=30$ achieved the lowest final loss, the performance gain was diminishing relative to the increase in training time. Additionally, the curve for $N_{\text{DP}}=40$ indicates unstable convergence behavior. These results confirm that 20 training samples provide sufficient information for stable learning while maintaining computational efficiency.

\section{Validation of spatio-temporal subsampling consistency}
\label{appendix:subsampling}
\renewcommand{\thefigure}{C.\arabic{figure}}
\renewcommand{\thetable}{C.\arabic{table}}
\setcounter{figure}{0}
\setcounter{table}{0}

We conducted an additional experiment to evaluate the effect of spatio-temporal subsampling on model convergence and generalization. All experiments used the unified model (coupled MLP) with a fixed training dataset of 20 data points, as validated in \ref{appendix:data_sensitivity}. The spatio-temporal sampling ratio, denoted as $r_{\text{sample}}$, was varied among $1\%$, $5\%$, $10\%$, $30\%$, and $100\%$ to investigate the influence of data reduction on model performance. Each configuration represents the proportion of total spatio-temporal data used per epoch.

Table~\ref{tab:appendix_samp_stats} summarizes the validation loss and computation time. Using the full $100\%$ dataset ($0.00225$) resulted in a higher validation loss than the $10\%$ ($0.00208$) and $30\%$ ($0.00199$) configurations. This result suggests that spatio-temporal subsampling improves generalization by preventing overfitting. The lowest loss ($0.00199$) was achieved at $r_{\text{sample}}=30\%$. However, increasing the ratio from $10\%$ to $30\%$ incurred an $84.5\%$ increase in computational cost (from $10.65\,\text{s}$ to $19.65\,\text{s}$) for only a $4.3\%$ reduction in loss. Therefore, considering generalization, stable convergence, and computational efficiency, we selected the $10\%$ subsampling ratio as the optimal configuration for subsequent experiments.
\begin{table}[H]
\centering
\caption{Validation loss and computation time per epoch for different spatio-temporal subsampling ratios.}
\label{tab:appendix_samp_stats}
\begin{tabular}{r r r}
\hline
\makecell{\textbf{$r_{\text{sample}}$ (\%)}} &
\makecell{\textbf{Best validation loss}} &
\makecell{\textbf{Time per epoch (s)}} \\ \hline
1   & 0.00231 &   5.26 \\
5   & 0.00215 &   6.67 \\
\textbf{10}  & \textbf{0.00208} & \textbf{10.65} \\
30  & 0.00199 &  19.65 \\
100 & 0.00225 &  60.48 \\ \hline
\end{tabular}
\end{table}

\begin{figure}[H]
    \centering
    \includegraphics[width=0.85\linewidth]{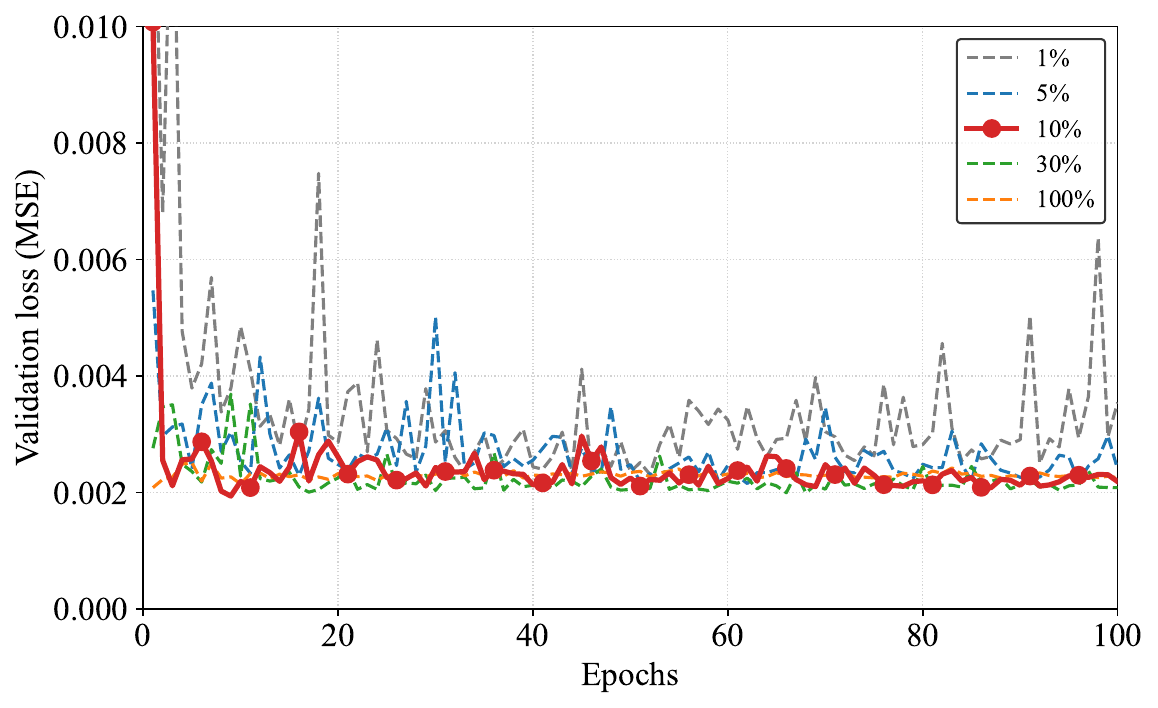}
    \caption{
    Validation loss curves with different spatio-temporal subsampling ratios ($r_{\text{sample}} = 1\%, 5\%, 10\%, 30\%, 100\%$). All experiments used the unified model (coupled MLP) with $20$ training data points. The $10\%$ and $30\%$ ratios achieved lower loss than the $100\%$ dataset, suggesting a regularization effect. The $10\%$ ratio was selected as the optimal trade-off, balancing generalization ($0.00208$ loss) and computational efficiency ($84.5\%$ faster per epoch than the $30\%$ ratio).
    }
    \label{fig:appendix_subsampling}
\end{figure}

Figure~\ref{fig:appendix_subsampling} shows the validation loss curves over $100$ epochs. When the sampling ratio was below $5\%$, the validation loss exhibited unstable fluctuations due to insufficient spatio-temporal coverage. In contrast, the $10\%$ and $30\%$ ratios achieved smooth convergence and lower final loss levels than the $100\%$ dataset. This result confirms that $10\%$ subsampling provides an effective trade-off, achieving robust generalization and stability while maintaining computational efficiency.

% \section{Model parameters and computational cost}
% \label{appendix:model_specs}
% [Detailed specifications and timing results]

\end{document}